\documentclass[a4paper,11pt]{article}
\usepackage{jheppub}

\usepackage[utf8]{inputenc}
\usepackage{amsmath}
\usepackage{amssymb}
\usepackage{graphicx}
\usepackage{multirow}
\usepackage{adjustbox} 

\usepackage{amssymb}
\usepackage{amsmath}
\usepackage{graphicx}
\def\vareps{\varepsilon}

\allowdisplaybreaks[1]

\numberwithin{equation}{section}

\DeclareMathOperator{\diag}{Diag}

\begin{document}

\thispagestyle{empty}
\begin{flushright}
    FERMILAB-PUB-24-0776-T\\

\vspace*{2.mm} \today
\end{flushright}

\title{Constraining non-standard neutrino interactions with neutral current events at long-baseline oscillation experiments}

\author[a]{Julia Gehrlein,}
\author[b]{Pedro A.~N.~Machado,}
\author[c]{and João Paulo
Pinheiro}

\affiliation[a]{Physics Department, Colorado State University, Fort Collins, CO 80523, USA}
\affiliation[b]{Theoretical Physics Department, Fermi National Accelerator Laboratory, Batavia, Illinois 60510}
\affiliation[c]{Departament de Física Quàntica i Astrofísica and Institut de Ciències del Cosmos, Universitat
de Barcelona, Diagonal 647, E-08028 Barcelona, Spain}

\emailAdd{julia.gehrlein@colostate.edu}
\emailAdd{pmachado@fnal.gov}
\emailAdd{joaopaulo.pinheiro@fqa.ub.edu}

\makeatletter
\hypersetup{colorlinks=true,allcolors=[rgb]{0.5,0.,0.5},pdftitle=\@title}
\makeatother

\abstract{
We explore, for the first time, {\textit{neutral-current}} events at long-baseline  experiments to constrain vector and axial-vector neutrino non-standard interactions (NSI) with quarks.
We leverage the flavor dependence of NSIs to perform an oscillation analysis in the neutral-current channel.
We first introduce a framework to parametrize the effect of NSI on the cross section.
Then, as an example, we analyze NOvA neutral-current data which 
provides significantly improved constraints on the axial-vector NSI parameters $\varepsilon_{\mu\mu}^A,~\varepsilon_{\tau\tau}^A$ and $\varepsilon_{e\mu}^A$. 
This is highly complementary to constraints from SNO data, which, differently from long-baseline neutral current data,  is not sensitive to isospin conserving NSIs  $\varepsilon^A_u=\varepsilon^A_d$.
Additionally, we disfavor large values of the diagonal vectorial NSI $\varepsilon_{\mu\mu}^V$ and $\varepsilon_{\tau\tau}^V$ which originate from the LMA-Dark solution. 
We also highlight the complementarity between NSI searches at oscillation experiments using charged current and neutral current channels.
}

\maketitle

\section{Introduction}
The observation of neutrino oscillations presents one of the few robust, laboratory-based evidences for  physics beyond the standard model (BSM).
Neutrinos being promising windows to new physics motivates a number of BSM searches in oscillation experiments.
Of particular interest are searches for nonstandard neutrino interactions (NSI), which are effective four-fermion interactions that can parametrize a number of BSM scenarios~\cite{Heeck:2011wj, Farzan:2015doa,Farzan:2015hkd, Farzan:2016wym,Forero:2016ghr,Babu:2017olk, Heeck:2018nzc,Denton:2018dqq,Dey:2018yht, Babu:2019mfe, Babu:2020nna, Babu:2021cxe}.
While new neutrino charged current interactions are already tightly constrained~\cite{Falkowski:2021bkq} neutral current (NC) interactions with interaction strength $\mathcal{O}(0.1~G_F)$, with $G_F$ the Fermi constant,  are still experimentally viable~\cite{Coloma:2023ixt}. 

There are two main probes of NC NSIs, via oscillation physics or neutrino NC scattering observables.
In the former case, since oscillations are observed by comparing the charged-current (CC) rate of events in near and far detectors, the only effect of (vector) NC NSIs are modifications to the matter potential.
This has been used to constrain NC NSI at a variety of long-baseline~\cite{NOvA:2024lti, MINOS:2013hmj}, atmospheric~\cite{IceCube:2017zcu,IceCubeCollaboration:2021euf,IceCube:2022ubv,Super-Kamiokande:2011dam} and solar neutrino experiments~\cite{Super-Kamiokande:2022lyl}. 
While oscillation physics provides leading constraints on vector NC NSIs, it cannot probe universal, flavor-diagonal NSIs, as these induce an overall phase in the oscillation amplitude which cancels out in the probability level~\cite{Proceedings:2019qno}.
On top of that, since matter effects can be described as coherent forward scattering processes, they are fairly independent of the mass of the new physics  which leads to NSI.\footnote{An exception to this statement is the case where the wavelength of the mediator is of the order of the depth the neutrinos travel through the Earth which leads to a depth-dependent potential due to the new mediator
\cite{Grifols:2003gy,Joshipura:2003jh,Davoudiasl:2011sz,Wise:2018rnb,Smirnov:2019cae,Babu:2019iml,Coloma:2020gfv, Chauhan:2024qew}.}

A complementary way to probe NC NSIs can be found in neutral current neutrino scattering experiments~\cite{Dutta:2020che,Coloma:2022umy, Coloma:2023ixt} which can probe both vectorial and axial-vector NSI. 
Scattering experiments typically have a detector very close to the neutrino source, and can measure all NC NSI parameters, subject to the source flavor composition.
Nevertheless, since the observable is very simple, that is, the number of NC scattering events, there are large degeneracies among NSI parameters \cite{COHERENT:2017ipa,COHERENT:2020iec,COHERENT:2021xmm,Adamski:2024yqt,Colaresi:2022obx}.
Moreover, as the scattering involves a non-negligible momentum transfer, these constraints exhibit a stronger dependence on the mass of the NSI mediator, especially for masses below the GeV scale.

In this manuscript we introduce a new way to test NC NSI in long-baseline accelerator neutrino experiments, combining the effects on neutrino oscillations and scattering rates.
Our proposal is to use neutral current event samples in these oscillation experiments to probe NC NSIs, potentially breaking some of the degeneracies present in the other experimental setups.
In the presence of NC NSIs with quarks, the neutral current neutrino-nucleus interactions change in a flavor-dependent way, effectively distinguishing the neutrino flavors in the neutral current channel.
This flavor non-universality makes oscillation physics relevant in neutral current observables.
On top of that, there are nontrivial quantum interference effects involving oscillation amplitudes and flavor-changing scatterings, as we will see later.
So far, the NC data set has been used to search for sterile neutrinos at the MINOS and NOvA experiments~\cite{MINOS:2008ccf, NOvA:2021smv, NOvA:2024imi} and it has been pointed out that such studies at future experiments can improve over current bounds on sterile neutrinos~\cite{Gandhi:2017vzo, Coloma:2017ptb, Dutta:2019hmb}, but to our knowledge there has not been a use of existing NC data in long-baseline experiments to probe NSIs.

A previous work focused on a potential future DUNE-like experiment \cite{Abbaslu:2023vqk} and explored NC Deep Inelastic Scattering (DIS) events, deriving bounds on the axial NC NSI of neutrinos with the $u$, $d$, and $s$ quarks. We  provide a more comprehensive analysis in this paper. We use existing NOvA data from both near and far detectors (ND and FD) to reduce systematic uncertainties and consider all types of cross sections relevant at the typical NOvA energies.

We will show that probing NC NSIs in long-baseline experiments with neutral current events is complementary to the usual searches, namely, NC NSI searches with long-baseline charged current event samples or short-baseline scattering events.
As an example, we will focus on the NOvA experiment and derive 
constraints from the existing NC data set using $13.6\times 10^{20}$ proton-on-target \cite{NOvA:2024imi}. 
We will introduce a simple, approximate parametrization of the impact of NC NSIs in propagation and interactions simultaneously which is valid for all oscillation experiments.
Furthermore, we will show that the presence of a near detector is crucial to probe NC NSI within this setup.

The manuscript is organized as follows: in Sec.~\ref{sec:framework}
we introduce the framework to incorporate NC NSI in propagation and detection, in Sec.~\ref{sec:effects} we study the effects of NC NSI at the near and far detector of NOvA. We present the constraints on NSI from existing and simulated NOvA data in Sec.~\ref{sec:results} before we conclude in Sec.~\ref{sec:conclusions}. The appendix contains details on the analysis like the SM prediction for the NC cross section in Sec.~\ref{sec:smnc}
 and the migration matrix we use to analyze the NOvA data in Sec.~\ref{sec:migration}.
 
\section{Framework}
\label{sec:framework}
New neutrino interactions  can be parametrized in an effective field theory approach where they can originate, for example, from dimension-6 effective operators which couple four SM fermions to each other, two of which are neutrinos. 
We focus on NSIs which have flavor conserving or flavor-changing couplings to neutrinos but we constrain ourselves to flavor conserving couplings to charged fermions.
Concretely, below electroweak symmetry breaking, we have 
\begin{align}
&\mathcal L_{\rm NSI}=-2\sqrt2G_F\sum_{f,\alpha,\beta}(\bar\nu_\alpha\gamma^\mu P_L\nu_\beta)\left[\vareps^{f,V}_{\alpha\beta}(\bar f\gamma_\mu f) + \vareps^{f,A}_{\alpha\beta}(\bar f\gamma_\mu \gamma_5 f) \right]\,,
\end{align}
where $\alpha,\beta$ are the neutrino flavor indices, $f=[e,u,d]$ are the first generation charged fermions, $G_F$ is the Fermi constant and $\varepsilon^{f,V}_{\alpha\beta},\varepsilon^{f,A}_{\alpha\beta}$ parametrize the strength of the NSI relative to weak interactions for vector and axial-vector interactions with charged fermions. 
While diagonal NSIs ($\alpha=\beta$) are real, off-diagonal NSIs  ($\alpha\neq\beta$) can be complex.
Vectorial neutral current NSIs affect oscillation physics by inducing a nonstandard matter potential for neutrinos traversing usual matter. 
The Hamiltonian governing neutrino oscillations in the flavor basis reads
\begin{equation}
H=\frac1{2E}\left[U^\dagger
M^2
U+a
\begin{pmatrix}
1+\mathcal{E}_{ee}&\mathcal{E}_{e\mu}&\mathcal{E}_{e\tau}\\
\mathcal{E}^{*}_{e\mu}&\mathcal{E}_{\mu\mu}&\mathcal{E}_{\mu\tau}\\
\mathcal{E}^{*}_{e\tau}&\mathcal{E}^{*}_{\mu\tau}&\mathcal{E}_{\tau\tau}
\end{pmatrix}
\right]\,,
\label{eq:HNSI}
\end{equation}
where $E$ is the neutrino energy, 
$U$ 
is the PMNS mixing matrix \cite{Pontecorvo:1957cp,Maki:1962mu} and $M^2\equiv\diag(0,\Delta m^2_{21},\Delta m^2_{31})$ is the diagonal mass-squared matrix, $a\equiv2\sqrt2G_FN_eE$ parametrizes the matter effect, and $N_e$ is the electron density.
We also have defined 
\begin{equation}
    \mathcal{E}_{\alpha\beta}\equiv \sum_f^{e,p,n}\frac{N_f}{N_e}\varepsilon_{\alpha\beta}^{f,V},
\end{equation}
such that 
\begin{equation}
\varepsilon_{\alpha\beta}^p=2\varepsilon_{\alpha\beta}^u + \varepsilon_{\alpha\beta}^d \,\,\,\mathrm{and}\,\,\,\varepsilon_{\alpha\beta}^n=\varepsilon_{\alpha\beta}^u + 2\varepsilon_{\alpha\beta}^d.
    \label{definition_epsilon}
\end{equation}
To simultaneously incorporate the effects of NSI in both  propagation and  interactions, we follow the approach outlined in Ref.~\cite{Coloma:2022umy}. 
In this framework, the number of events is proportional to the trace
\begin{equation}\label{eq:trace}
    N_{\mathrm{ev}} \propto \sum_\gamma\text{tr}(\rho \hat\sigma_\gamma),
\end{equation} 
where $\rho$ is the density matrix describing the neutrino system, and $\hat\sigma_\gamma$ is a generalized cross section operator which will be described later.
The trace is applied to the incoming neutrino flavors, while the different final state neutrino flavor $\gamma$ is summed up incoherently.

The density matrix is calculated as usual, namely $\rho(L) = \exp^{-i H L} \rho(0) \exp^{i H L}$, where $H$ is the Hamiltonian describing neutrino oscillations in matter in the presence of NSI, as given by Eq.~(\ref{eq:HNSI}). 
For the NOvA experiment, we will consider the forward horn configuration only, in which the flux is dominated by muon neutrinos~\cite{NOvA:2004blv}.
Therefore, we approximate the initial density matrix at the near detector, $L= 0$, by
\[
\rho(0) = \begin{pmatrix}
0 & 0 & 0 \\
0 & 1 & 0 \\
0 & 0 & 0
\end{pmatrix}.
\]
Now let us consider the generalized cross section operator to incorporate the effects of NSI on interactions following Ref.~\cite{Coloma:2022umy, Amaral:2023tbs}.
We will start by considering the elastic scattering of neutrinos off free nucleons at rest.
The generalized cross section operator per nucleon $N=p,n$, for any incoming neutrino flavor, is given by~\footnote{
    If $\hat T=\sum_a c_a |\gamma\rangle\langle \nu_a|$ describes the transition from an incoming neutrino state $|\nu\rangle$ to an outgoing state $|\gamma\rangle$, we can apply the transition operator to a incoming state and obtain $|\psi\rangle=\hat T |\nu\rangle$. Therefore $|\langle \psi|\psi\rangle|^2 = {\rm Tr} (\rho T^\dagger T) \propto {\rm Tr} (\rho \hat\sigma_\gamma)$, where $\rho=|\nu\rangle\langle\nu|$. This can be trivially generalized to mixed states.}
\begin{equation}
    \hat\sigma(\nu N\to \nu_\gamma N) = \sum_{\alpha,\beta} |\nu_\alpha\rangle \sigma_{N\gamma}^{\alpha\beta}\langle \nu_\beta|.
\end{equation} 
In the standard model , $\sigma_{N\gamma}^{\alpha\beta}\neq0$ only if $\alpha = \beta = \gamma$.
The differential generalized cross section reads
\begin{equation}
 \dfrac{d \sigma^{\alpha\beta}_{N\gamma}}{d T_N}= \dfrac{2 G_F^2 m_N}{\pi} \!\left[ G_{\gamma \alpha}^{N\!L\star}\,G_{\gamma \beta}^{N\!L} 
 + G_{\gamma \alpha}^{N\!R\star}\,G_{\gamma \beta}^{N\!R}\left(1-\frac{T_N}{E_\nu}\right)^2 \!\!
 -  \left( G_{\gamma \beta}^{N\!R} G_{\gamma \alpha}^{N\!L\star}+G_{\gamma \alpha}^{N\!R\star}G_{\gamma \beta}^{N\!L}\right)\dfrac{m_N T_N}{2 E_\nu^2}\right]\!,
    \label{eq:genxsec}
\end{equation}
where $T_N$ is the recoil energy of the nucleon $N$.
We define the generalized couplings
\begin{eqnarray}
 G_{\gamma \alpha}^{N\!L} = g_L^N\delta_{\gamma\alpha} + \varepsilon_{\gamma\alpha}^{N\!L}, \\
 G_{\gamma \alpha}^{N\!R} = g_R^N\delta_{\gamma\alpha} + \varepsilon_{\gamma \alpha}^{N\!R},
\end{eqnarray}
where we have defined $\varepsilon_{\gamma\alpha}^{N\!R} = (\vareps_{\gamma\alpha}^{NV} + \vareps_{\gamma\alpha}^{N\!A})/2$ and $\vareps_{\gamma\alpha}^{N\!L} = (\vareps_{\gamma\alpha}^{NV} - \vareps_{\gamma\alpha}^{N\!A})/2$, and the nucleon NSI parameter is the sum of the valence quark NSIs (e.g. $\vareps^p=2\vareps^u+\vareps^d$). For the standard $Z$-boson couplings, we have
\begin{eqnarray}
  g_{L}^p=\frac{1}{2}-s_W^2,& \qquad &g_{R}^p=-s_W^2,\nonumber\\
  g_{L}^n=-\frac{1}{2},& \qquad  &g_{R}^n=0,
\end{eqnarray}
where $s_W^2=0.23$ is the weak mixing angle.
Again, we highlight that the indices $\alpha,~\beta$ refer to the incoming neutrino flavor, while $\gamma$ corresponds to the outgoing neutrino flavor.
The off-diagonal elements of the generalized cross section  (that is, when $\alpha\neq\beta$) parametrize the interference arising from the fact that the incoming neutrino state is a coherent superposition of flavor states.
Since the final state neutrino is not observable, we will always be summing up on the outgoing neutrino flavor, 
\begin{equation}
  \dfrac{d \sigma^{\alpha\beta}_{N}}{d T_N} \equiv \sum_\gamma^{e,\mu,\tau}\dfrac{d \sigma^{\alpha\beta}_{N\gamma}}{d T_N},
\end{equation}
and similarly for the total generalized cross section.

The number of events for neutrinos scattering off nucleons, in the presence of NSIs, is proportional to
\begin{equation}  
N_\mathrm{ev}^\mathrm{free} \propto \sum_{\alpha,\beta}^{e,\mu,\tau} \left(N_p \rho_{\alpha\beta} \sigma^{\alpha\beta}_{p} + N_n \rho_{\alpha\beta} \sigma^{\alpha\beta}_{n}\right),
\label{eq:xsec1}
\end{equation}
where $N_p,\, N_n$ represent the number of free protons and neutrons.
It is important to note that the formalism above specifically applies to the simplified scenario of free nucleons at rest, without internal structure. 
A more precise treatment of NSI effects in the context of coherent elastic neutrino-nucleus scattering is provided in Ref.~\cite{Coloma:2023ixt}.

However, the energy spectra in accelerator neutrino experiments typically range from about 0.3 to 10 GeV, a region in which nuclear and nonperturbative effects are relevant (see App.~\ref{sec:smnc}). 
To properly account for these, the generalized cross section should be interfaced with the hadronic current in the description of neutrino-nucleus interactions.
While a detailed interface within the framework of the \texttt{Achilles} event generator~\cite{Isaacson:2020wlx, Isaacson:2022cwh} will be left to future work,  we propose here the following simplified approach.
We use \texttt{NuWro}~\cite{Juszczak:2005zs,Golan:2012wx,Golan:2012rfa}  to generate standard neutrino neutral current events in the detector material, which for NOvA is CH$_2$. 
Then we approximate the NC events in the presence of NSIs by 
\begin{equation}  
N_\mathrm{ev} \propto  \frac{\sum_{\alpha\beta} N_p \rho_{\alpha\beta} \sigma^{\alpha\beta}_p + N_n \rho_{\alpha\beta} \sigma^{\alpha\beta}_n}{\sum_{\alpha} N_p {\rho_{\alpha\alpha} (\sigma^{\alpha\alpha}_p})_{\rm SM} + N_n {\rho_{\alpha\alpha} (\sigma^{\alpha\alpha}_n})_{\rm SM}} \sigma_{\mathrm{NuWro}},
\label{eq:xsec}
\end{equation}
where $\sigma_{\mathrm{NuWro}}$ is the standard NC cross section from \texttt{NuWro}~\footnote{Appendix~\ref{sec:smnc} details the different contributions to the NC cross section, with quasi-elastic and resonant scattering dominating near the peak of the neutrino flux in NOvA.}, while $\sigma_{p,n}^{\alpha\beta}$ and $(\sigma_{p,n}^{\alpha\beta})_{\rm SM}$ are the generalized and standard NC cross sections on free protons and neutrons at rest.
In practice, the prefactor can be used to reweight \texttt{NuWro} events to account for the impact of NSIs. 

Our approach trivially reduces to the standard cross section when there are no NSIs present, as the prefactor cancels out and $N_\mathrm{ev} \propto \sigma_{\mathrm{NuWro}}$.
For the case of free nucleons at rest, Eq.~(\ref{eq:xsec}) should, in principle, be exact, although we are simplifying the nucleon as a point particle in this first work. Nevertheless, since we are only considering vector and axial-vector NSIs, we expect this approximation to be reasonably accurate (see Ref.~\cite{Kopp:2024yvh}). However, note that  deviations from our approach are expected when taking form factors into account for example due to the
specific flavor dependence of  various inelastic processes, from the phase space of a final state with multiple particles, and the effect of strange quarks in the nucleon \cite{Ilma:2024lkp}. 

In \cite{Ilma:2024lkp} it has been shown that in the presence of axial NSI the elastic NC cross section depends on the isoscalar form factor of the nucleon which is not probed in SM processes, demonstrating that the inclusion of form factors in this analysis will further probe our understanding of nuclear effects, beyond the present study.

Working within this approximation but at the event level will allow us to better estimate the deposited energy in the neutral current sample.

The expression for the number of observed events
 is  given by 
\begin{eqnarray}
   &&N_{\mathrm{ev}}(E_{\rm rec}) =\Xi \dfrac{d\phi(E_\nu)}{dE_\nu}  
\times\frac{\sum_{\alpha\beta}N_p\rho_{\alpha\beta} \sigma^{\alpha\beta}_p + N_n\rho_{\alpha\beta} \sigma^{\alpha\beta}_n}{\sum_{\alpha}N_p({\rho_{\alpha\alpha} \sigma^{\alpha\alpha}_p})_{\rm SM} + N_n({\rho_{\alpha\alpha} \sigma^{\alpha\alpha}_n})_{\rm SM}} \sigma_{\mathrm{NuWro}}(E_\nu) \times R(E_\nu,~E_{\rm rec}),\nonumber
\end{eqnarray}
where $\phi(E_\nu)$ is the initial neutrino flux as a function of the neutrino energy, $\Xi = N_{\rm tgt} T$   encodes the number of target nuclei $N_{\rm tgt}$ in a detector and the exposure $T$ (scaled by the distance to the detector), and $R$ is a migration matrix that maps a neutrino energy into a reconstructed energy.
Since we focus on the NOvA experiment, we have derived $R$ via an event-by-event analysis of NOvA's detector response, see App.~\ref{sec:migration}. 
This can easily be generalized to other neutrino experiments such as MINOS and DUNE.
Last, we note that the above expression for the NSI interaction is only valid as an effective operator for scales above the momentum transfer in the scattering. 
In the case of NOvA, the momentum transfer spectrum peaks at 0.3~GeV and rapidly falls for higher values.

\section{Effect of NC NSI at NOvA}
\label{sec:effects}

Let us study the effect of NC NSI in long-baseline experiments with a concrete example by analyzing the  recent NOvA NC data~\cite{NOvA:2024imi}.
The NOvA experiment~\cite{NOvA:2004blv} consists of two detectors downstream the Neutrinos at the Main Injector (NuMI) beam at Fermilab: a 193 ton near detector located 1~km away from the target, and a 14~kton far detector 810~km further. 
The neutrino beam spans an energy spectrum up to about 20~GeV, with its peak around 2 GeV which rapidly drops at higher neutrino energies. 
Ref~\cite{NOvA:2024imi} analyzes an exposure of 13.6$\times 10^{20}$ protons-on-target in the forward horn configuration mode.

The key novelty of using NC events in long-baseline experiments to probe NC NSIs lies in the interplay between oscillation and scattering physics.
In the standard model, NC interactions do not distinguish flavor, and since the final state neutrino flavor is not detected oscillation effects are unobservable.
Here, the NC NSI are flavor dependent, making oscillations relevant in the neutral current sample. 
In order to simplify our analysis, we are assuming isospin conserving NSIs,
\begin{equation}
\epsilon^{V(A)}_{\alpha\beta}=\epsilon^{pV(A)}_{\alpha\beta}=\epsilon^{nV(A)}_{\alpha\beta}.
    \label{approach_nucleon_nsi}
\end{equation}

\begin{figure}[t]
    \centering
\includegraphics[scale=0.4]{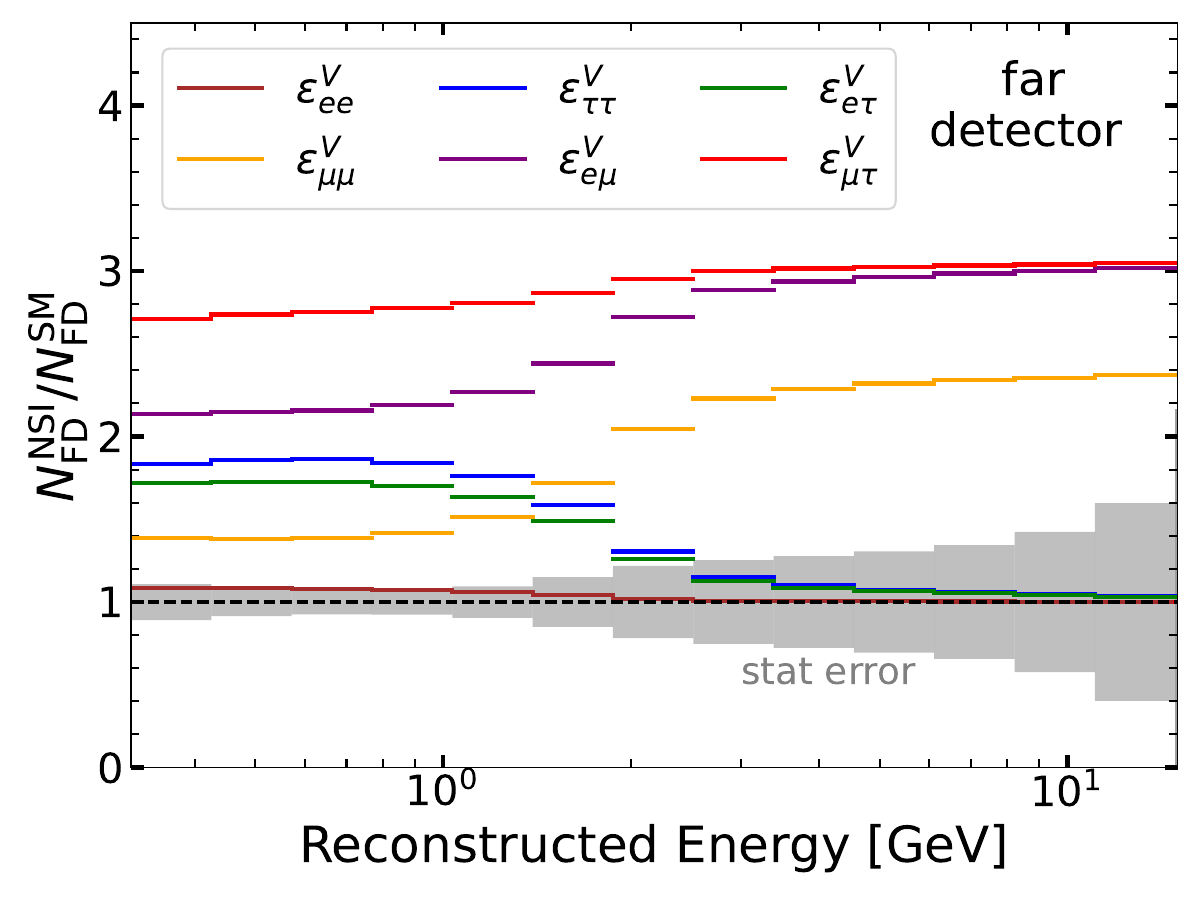}
    \caption{Ratio of the number of events at the NOvA far detector in the presence of NSI divided by the SM prediction at the far detector as a function of the reconstructed energy. We consider only one vectorial NSI parameter at a time which we fix to $\varepsilon^V_{\alpha\beta}=1$. For comparison we show the statistical error bar on the number of events at the far detector from Ref.~\cite{NOvA:2024imi} as a gray band.}
    \label{fig:1}
\end{figure}

To demonstrate the interplay between NC NSI and oscillations, we show in Fig.~\ref{fig:1} the number of NC events at NOvA's far detector (FD) in the presence of NC NSIs relative to the standard prediction, that is, $N_{\text{FD}}^{\text{NSI}}/N_{\text{FD}}^{\text{SM}}$.
For simplicity, we take only vectorial NSIs, but a similar phenomenology follows for axial-vector NSIs as we find that the dominant effect of NSIs in NOvA comes from scattering and not from the change in the matter potential.
We present six benchmarks, $\varepsilon^V_{\alpha\beta}= 1$, corresponding to each real NSI parameter taken to be unity while all others are set to zero.
Although some of these values are excluded by global fits~\cite{Coloma:2023ixt}, this figure serves to illustrate the  impact of  individual NSI parameters.
For NSIs this large, the number of events is always enhanced with respect to the standard scenario.

We can see that the increase in the number of events due to NSIs presents an energy dependence which is simple to understand.
At the far detector, a significant fraction of the dominant muon neutrinos have oscillated to tau neutrinos.
More concretely, at the far detector, the flavor ratio is approximately $\nu_e:\nu_\mu:\nu_\tau=1:11:14$.
The peak of oscillations happens at around $E_\nu=2$~GeV, which would correspond to reconstructed energies below about 1 GeV.
At peak, the flux is dominated by tau neutrinos.
At larger energies, oscillations do not have time to develop and the beam is dominated by muon neutrinos.
Therefore, NSIs that enhance the $\nu_\mu$ NC cross section such as $\vareps^V_{\mu\mu}$ and $\vareps^V_{e\mu}$, lead to a larger increase at higher energies. 
In contrast, $\vareps^V_{\tau\tau}$ and $\vareps^V_{e\tau}$ lead to larger increases of the number of events at lower energies, where tau neutrinos are more abundant.
For $\vareps^V_{\mu\tau}$, there is a large increase in the number of events, though somewhat constant in energy, since most $\nu_\mu$ oscillated to $\nu_\tau$. 
Last, since the $\nu_e$ contamination is very low, the impact of $\vareps^V_{ee}$ is fairly small.

\begin{figure}
    \centering
        \includegraphics[scale=0.35]{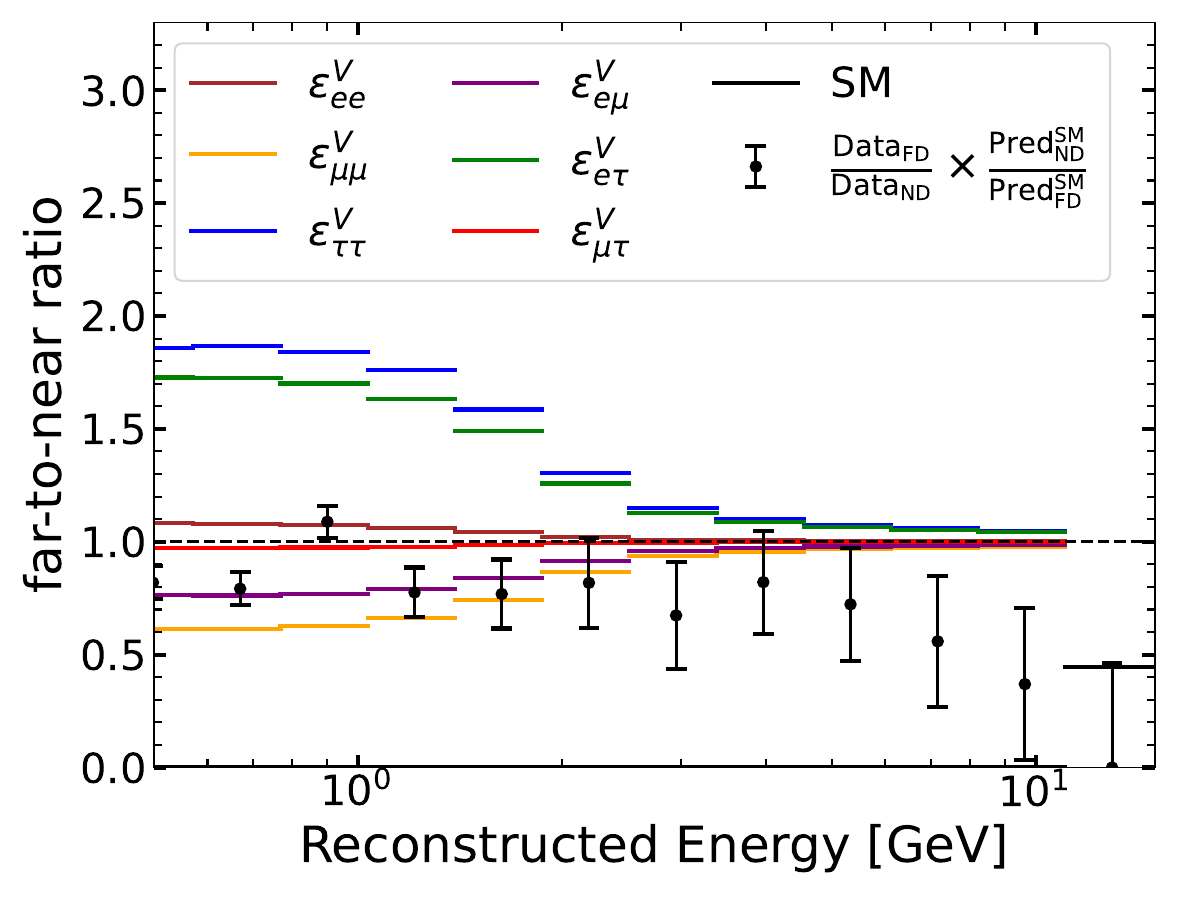}
        \includegraphics[scale=0.35]{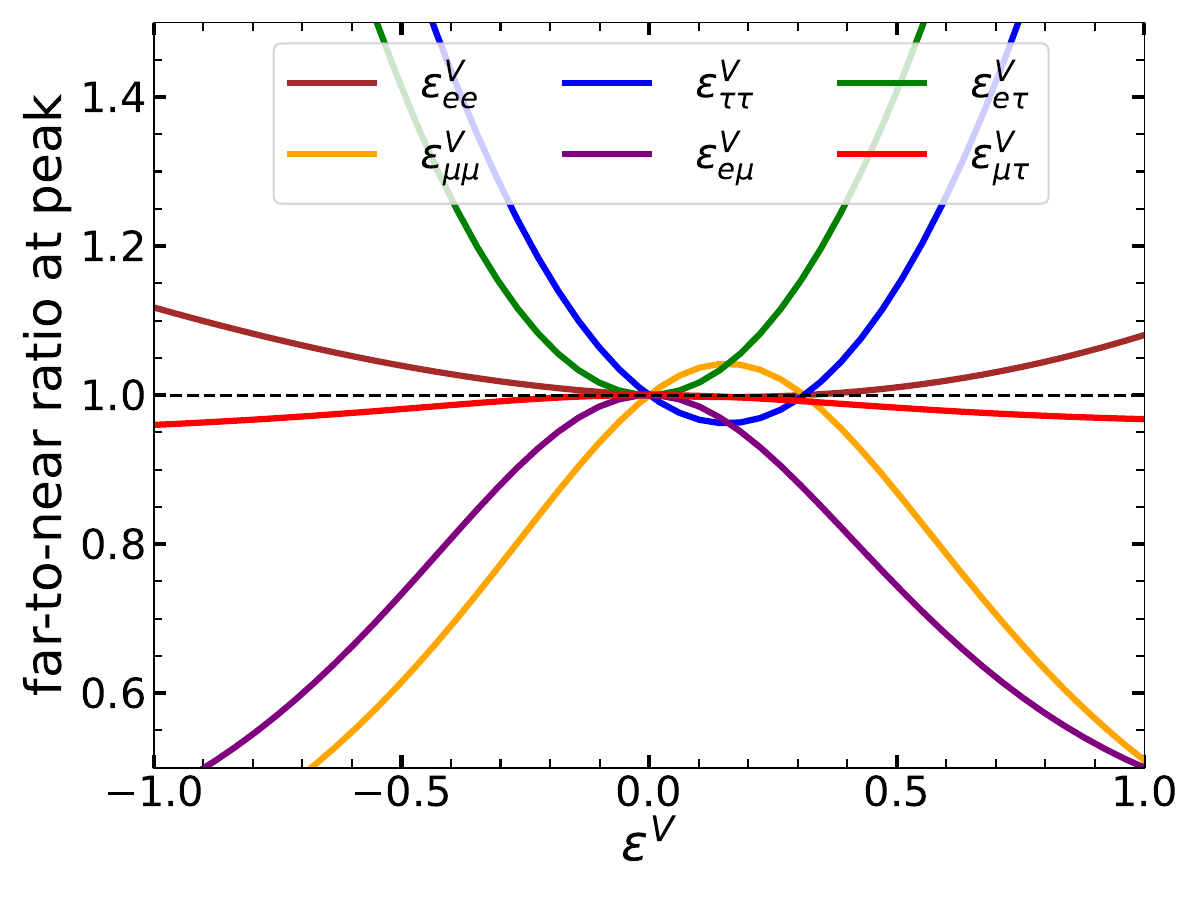}
    \caption{
    Left: Far-to-near detector event ratio for several NSI parameters normalized to the standard far-to-near detector event ratio prediction ($(N_{\text{FD}}^{\text{NSI}}/N_{\text{ND}}^{\text{NSI}})/(N_{\text{FD}}^{\text{SM}}/N_{\text{ND}}^{\text{SM}})$) as a function of the reconstructed energy. 
    We consider only one non-zero vectorial NSI parameter at a time which we fix to 1. 
    We also show the double ratio for the NOvA data with statistical uncertainties~\cite{NOvA:2024imi}.
    Right: Same far-to-near ratio, for fixed reconstructed energy $E_{\rm rec}=0.7$~GeV, as a function of the NC NSI parameter, taking only one nonzero parameter at a time.
    }
    \label{fig:fd-nd-ratios}
\end{figure}

While the far detector event rate help us understand the interplay between the effects of  NC NSIs in detection and in oscillations, a better approximation of a real experimental analysis is to look at the impact of NSIs on both far detector and near detector (ND) at the same time.
This is because there are significant systematic uncertainties related to the modeling of the neutrino flux and cross sections which tend to cancel when taking the ratio of far-to-near detector events.
We show in the left panel of Fig.~\ref{fig:fd-nd-ratios} the far-to-near ratio of events for several NSI benchmarks, $\vareps^V_{\alpha\beta}=1$ (one at a time),  normalized to the ratio in the SM, that is $(N_{\text{FD}}^{\text{NSI}}/N_{\text{ND}}^{\text{NSI}})/(N_{\text{FD}}^{\text{SM}}/N_{\text{ND}}^{\text{SM}})$, as a function of the reconstructed neutrino energy.
We also show NOvA's data with statistical uncertainties for comparison~\cite{NOvA:2024imi}.
As we can see, the far-to-near ratio can either increase or decrease, depending on the NSI parameter at play.
For $\vareps^V_{\mu\mu}$ and $\vareps^V_{e\mu}$, the enhancement of near detector (ND)  events due to NC NSIs  is larger than in the FD, simply due to the fact that in the far detector a significant fraction of the $\nu_\mu$ flux has oscillated away.
For NSIs that enhance the $\nu_\tau$ cross section, the behavior is the opposite.
Again, $\vareps^V_{\mu\tau}$ increase both $\nu_\mu$ and $\nu_\tau$ cross sections  and $\vareps^V_{ee}$ only change the small $\nu_e$ flux, both leading to small effects on the far-to-near ratio.

In the right panel of Fig.~\ref{fig:fd-nd-ratios}, we present the same far-to-near ratio for a representative value of the reconstructed energy $E_{\rm rec}=0.7$~GeV, as a function of the NC NSI parameter. 
Again, we take only one $\varepsilon^V_{\alpha\beta}$ to be nonzero at a time.
We can see a similar behavior regarding the impact of NSIs on the far-to-near ratio, as discussed above, and we can also see potential degeneracies between NSI parameters.
These degeneracies are physical and can be understood by performing the following approximations to Eqs.~(\ref{eq:trace}) and (\ref{eq:genxsec}).

First, we will consider a pure $\nu_\mu$ beam and a simplified two-neutrino framework in which $\nu_\mu$ only oscillates to $\nu_\tau$, such that $P_{\mu\mu}+P_{\mu\tau}=1$.
Then, we approximate $\sin^2\theta_W-1/4\simeq0$, and we take the same number of protons and neutrons in the target.
It is straightforward to show that, for $\vareps^V_{\mu\mu}$ and $\vareps^V_{\tau\tau}$, 
\begin{equation}
    \frac{{\rm Tr}(\rho\hat\sigma)_{\rm FD}}{{\rm Tr}(\rho\hat\sigma)_{\rm ND}} = 1 + (\vareps^V_{\mu\mu}-\vareps^V_{\tau\tau})(1-4\vareps^V_{\mu\mu}-4\vareps^V_{\tau\tau})\times(\dots).
\end{equation}
Therefore, there are two degenerate solutions that lead to no impact on the far-to-near ratio
\begin{equation}\label{eq:degeneracies}
    \vareps^V_{\tau\tau}=\vareps^V_{\mu\mu}\qquad{\rm and}\qquad
    \vareps^V_{\tau\tau}=\frac14-\vareps^V_{\mu\mu}.
\end{equation}
The first degeneracy originates in the fact that changing the NC cross section of all flavors by the same amount does not change the near-to-far ratio.
The second degeneracy is not present in the CC detection channel, but on the flip side, the NC channel does not suffer from the LMA-Dark~\cite{deGouvea:2000pqg,Miranda:2004nb,Bakhti:2014pva,Coloma:2016gei} degeneracy, offering a promising venue to solve that issue.
For the case of axial-vector NSIs, we obviously have a degeneracy when all diagonal NSIs have the same value, as is the case for the vector NSI.
Besides, since $g_A^p=1/2=-g_A^n$, the  differential cross section is symmetric under $\varepsilon^A_{\alpha\alpha}\to-\varepsilon^A_{\alpha\alpha}$ for any flavor. 
Therefore, we have, within our approximations
\begin{equation}\label{eq:axial-degeneracies}
    \varepsilon^A_{\tau\tau} = \varepsilon^A_{\mu\mu} \qquad{\rm and}\qquad
    \vareps^A_{\tau\tau}=-\vareps^A_{\mu\mu}.
\end{equation}

These degeneracies within vector and axial NSIs are very robust, as they are fairly independent of the neutrino energy. 
The dependence on the neutrino energy only comes from terms proportional to $\sin^2 2\theta_{13}\simeq0.1$, where $\theta_{13}$ is the smallest leptonic mixing angle, due to $\nu_\mu\to\nu_e$ oscillations; $1/4-\sin^2\theta_W\simeq0.02$; and the proton fraction of the target $1/2-Z/A$, which for NOvA's CH$_2$ would be about 0.07.
The left-hand degeneracies in Eqs.~(\ref{eq:degeneracies}) and (\ref{eq:axial-degeneracies}) are in fact universal changes to neutral current interactions, which become, when considering three neutrino flavors \begin{equation}
    \vareps^V_{ee}=\vareps^V_{\mu\mu}=\vareps^V_{\tau\tau} \quad{\rm and }\quad\vareps^A_{ee}=\vareps^A_{\mu\mu}=\vareps^A_{\tau\tau},
\end{equation}
with all other NSI parameters set to zero.
In this case, the NC NSIs do not distinguish flavors and the degeneracy is exact and independent of neutrino energy for both vector and axial NSIs. 
Note that in this case the overall ratio of events in both near and far detectors will be modified, but that is highly degenerate with flux and cross section uncertainties.
The only way to break this degeneracy is by measuring precisely the neutrino-nucleus neutral current interaction.

\section{Results}
\label{sec:results}
Having shown the impact of NC NSI on the number of events at long-baseline  experiments we now derive constraints on the NSI parameters. 
We analyze the NOvA data set with 13.6$\times 10^{20}$ POT from \cite{NOvA:2024imi}.
We assume that the neutrino mass ordering is normal and we use the best fit values of the oscillation parameters from \cite{NOvA:2024imi} which we keep fixed. A different choice for the true values of oscillation parameters will not affect our results much. We perform a ND and FD fit to the data and use a
 $\chi^2$-statistic 
\begin{align}
    \chi^2=\sum_i^ {\rm bins} \left[\frac{D_{i}^{\rm ND}-(1+\gamma_i)T_{i}^{\rm ND}}{\sqrt{D_{i}^{\rm ND}}}\right]^2
    +\sum_i^{\rm bins}&\left[\frac{D_{i}^{\rm FD}-(1+\alpha+\gamma_i)T_{i}^{\rm FD}}{\sqrt{D_{i}^{\rm FD}}}\right]^2 
    +\left(\frac{\alpha}{\sigma_\alpha}\right)^2
\end{align}
where $D$ and $T$ refer to `data' (either data or mock data assuming no NSIs) and `theory' (fit with NSIs). 
To be conservative, we include bin-to-bin pull parameters $\gamma_i$, fully correlated between the near and far detector, without any penalty term associated to it.
This will cover the large overall uncertainties on flux and NC cross section and possible near detector tunings performed in the analysis.
We also include a separate FD systematic uncertainty $\alpha$ with $\sigma_\alpha=0.15$, uncorrelated with the ND. 
Note that with all these systematics, NOvA's ND data can be fitted perfectly. 
We use a total of 14 reconstructed energy bins for the ND and FD between 300 MeV and 20 GeV, equally spaced in logarithmic scale.

\begin{figure}[t]
    \centering
        \begin{adjustbox}{valign=c} 
        \includegraphics[width=0.425\linewidth]{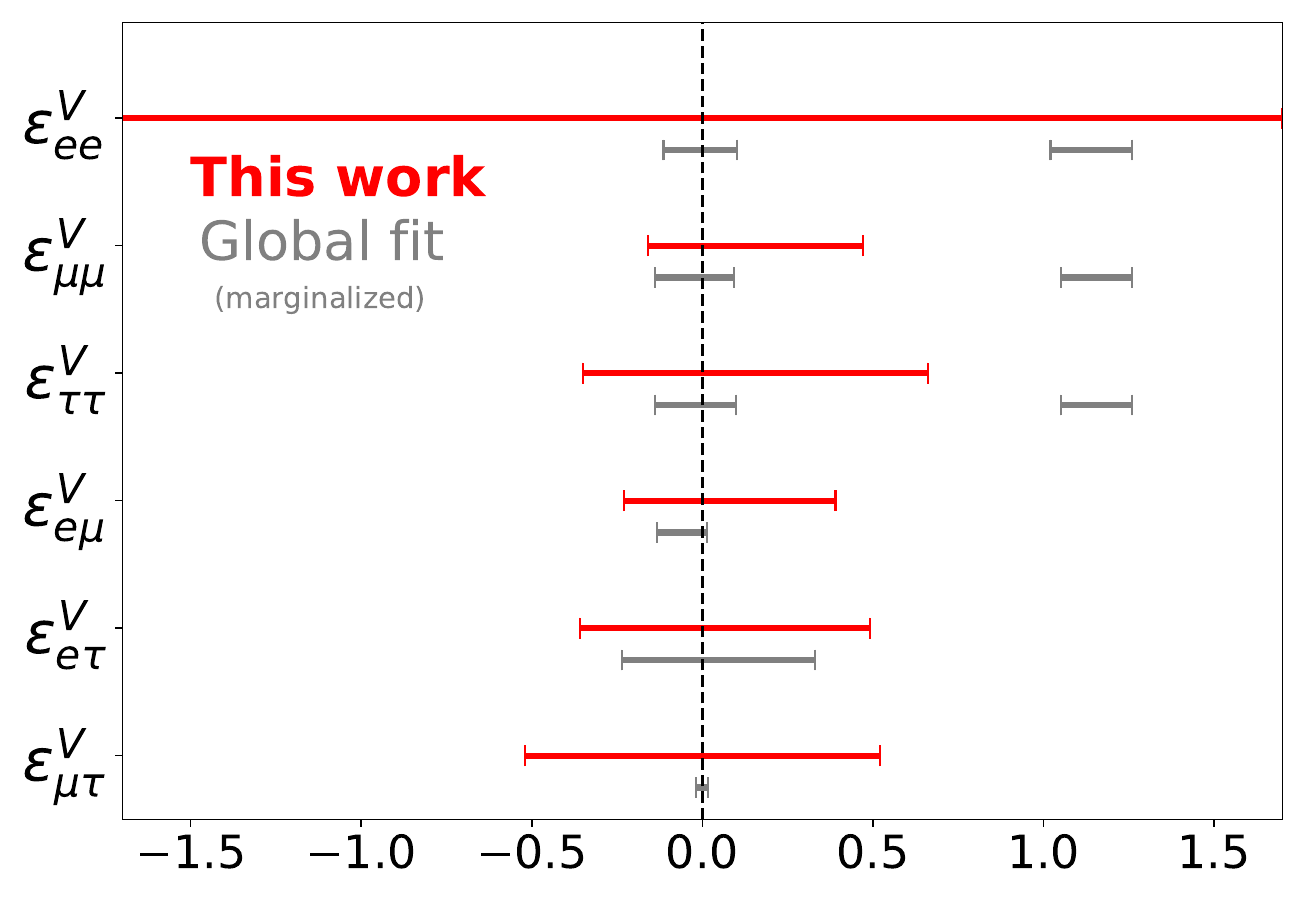}
        \end{adjustbox}
        \begin{adjustbox}{valign=c, raise=1.5mm} 
        \includegraphics[width=0.565\linewidth]{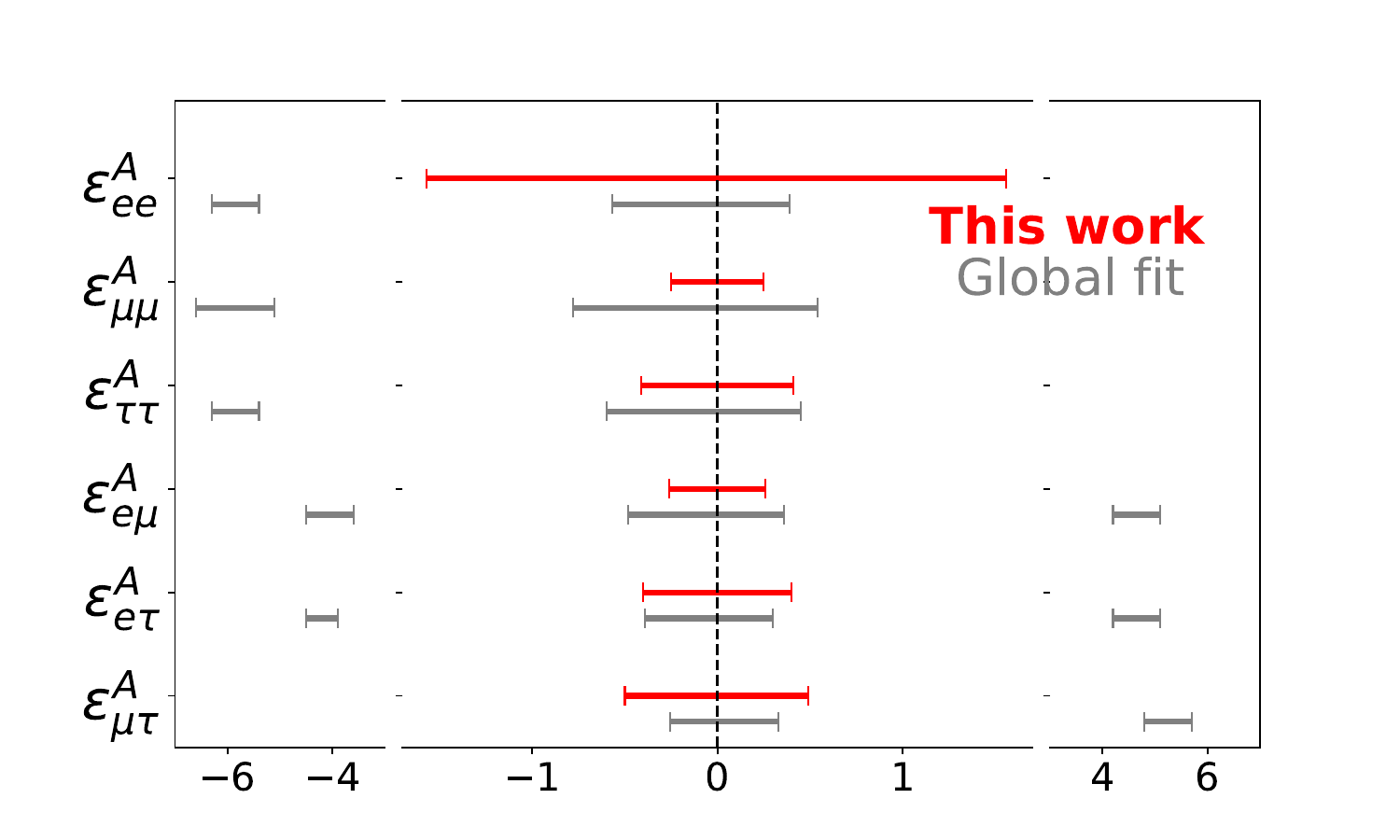}
        \end{adjustbox}
    \caption{We show in red the constraints on diagonal and off-diagonal vector (left panel) and axial (right panel) NC NSI parameters for actual NOvA data at 90$\%$ C.L. for an exposure of $13.6\times 10^{20}$ POT \cite{NOvA:2024imi}.  For comparison we show  the  90\% C.L.  global constraints from  scattering and oscillation data  in gray~\cite{Coloma:2023ixt}. For the sake of comparison, we are benchmarking our NSIs against the bounds given by \(3 \epsilon_{\alpha \beta}^{u(A,V)}\) of the global fit.
    We allow only one non-zero NSI parameter at a time and treat the off-diagonal parameters to be real.
    Note that the global constraints for vector NSIs have been derived by marginalizing in all parameters, while the ones for axial NSIs assume only one nonzero NSI at a time. See text for details.
    }
    \label{fig:nsi-1d-constraint}
\end{figure}

Turning now to the results, in Fig.~\ref{fig:nsi-1d-constraint}, we show the NOvA constraints to non-standard interactions  for the vectorial (left) and axial (right) NSI parameters, assuming one NSI parameter is non-zero at a time. 
For comparison, we also present the respective results from the global fit~\cite{Coloma:2023ixt}. 
To allow a comparison with the global fit results we rescale the global fit results bounds as $\epsilon_{\alpha\beta}^{(A,V)}=3\epsilon_{\alpha\beta}^{uA (uV)}$.  For the vectorial case, the global fit provides marginalized results where all NSI parameters are allowed to vary simultaneously. Although this hinders a direct comparison with our results and the global fit, our choice of  considering only one non-zero NSI parameter-at-a-time is motivated due to several degeneracies in the NSI parameter space which make it challenging for one experiment to derive constraints on one NSI parameter allowing all of them to be non-zero. Note that global fits overcome this challenge by considering several experiments simultaneously to break these degeneracies.
The inclusion of  NC NSI detection data, as proposed in this manuscript, can provide a valuable opportunity to break remaining degeneracies in the global fit. However a quantitative estimate of this is beyond the scope of this work. In contrast, for axial NSIs, the global fit show results obtained by fitting with only one non-zero parameter at a time.

%

Focusing on the vectorial case first, we observe that the global fit constrains the region near zero for all NSI parameters more tightly than current NOvA data. 
However, there are regions of large NSI values, particularly for $\varepsilon^V_{e\mu}$, $\varepsilon^V_{\mu\mu}$, and $\varepsilon^V_{\tau\tau}$ values near one, which originate from the LMA-Dark region~\cite{Coloma:2016gei}, where the constraints from the global fit are weaker. 
Our one-parameter-at-a-time fit excludes these regions, potentially contributing valuable information to the global fit.
The dominance of the global fit is not unexpected. 
Since vectorial NSIs change matter effects, therefore modifying oscillation probabilities, there are a large set of experiments that are sensitive to those new physics effects, and there are significant synergies among those experiments. However, only oscillation cannot break the degeneracy of the LMA-Dark solution. This is the role of COHERENT~\cite{COHERENT:2020iec, COHERENT:2021xmm} and DRESDEN-II \cite{Colaresi:2021kus}. In particular, since these experiments use neutrinos from pion decay at rest, muon three-body decay or from reactors, they are only sensitive to NSI affecting $\nu_e$ and $\nu_\mu$. And this is how LBL oscillation experiments are complementary to global fits. 

In the axial case, our results are more noteworthy. 
The NOvA neutral current analysis significantly improves constraints on the NSI parameters $\varepsilon_{\mu\mu}^A$, $\varepsilon_{\tau\tau}^A$, and $\varepsilon_{e\mu}^A$ compared to the global fit, and also helps break degeneracies for $\varepsilon_{e\tau}^A$ and $\varepsilon_{\mu\tau}^A$. 
The only parameter where we do not contribute significantly is $\varepsilon_{ee}^A$, although we do rule out a region of large NSI values. 
Here, the global fit is less sensitive to axial NSIs, as these do not affect oscillation probabilities but only NC cross sections (see also \cite{Coloma:2024ict}). The primary measurement used to probe axial NSIs with quarks comes from a single experiment, SNO \cite{SNO:2011hxd}, which is sensitive to the axial NC interactions of solar neutrinos with deuterium \cite{Bahcall:1988em} of the type $\varepsilon^A=\varepsilon^A_u-\varepsilon^A_d$. 
Therefore, incorporating neutral current channels in long-baseline experiments would not only improve the sensitivity to general axial NSIs, substantially strenghtening the constraints on these interactions, but also lead to exclusively new bounds for isospin conserving axial NSIs, $\varepsilon^A_u=\varepsilon^A_d$.

\begin{figure}[t]
    \centering
        \begin{adjustbox}{valign=c} 
        \includegraphics[scale=0.33]{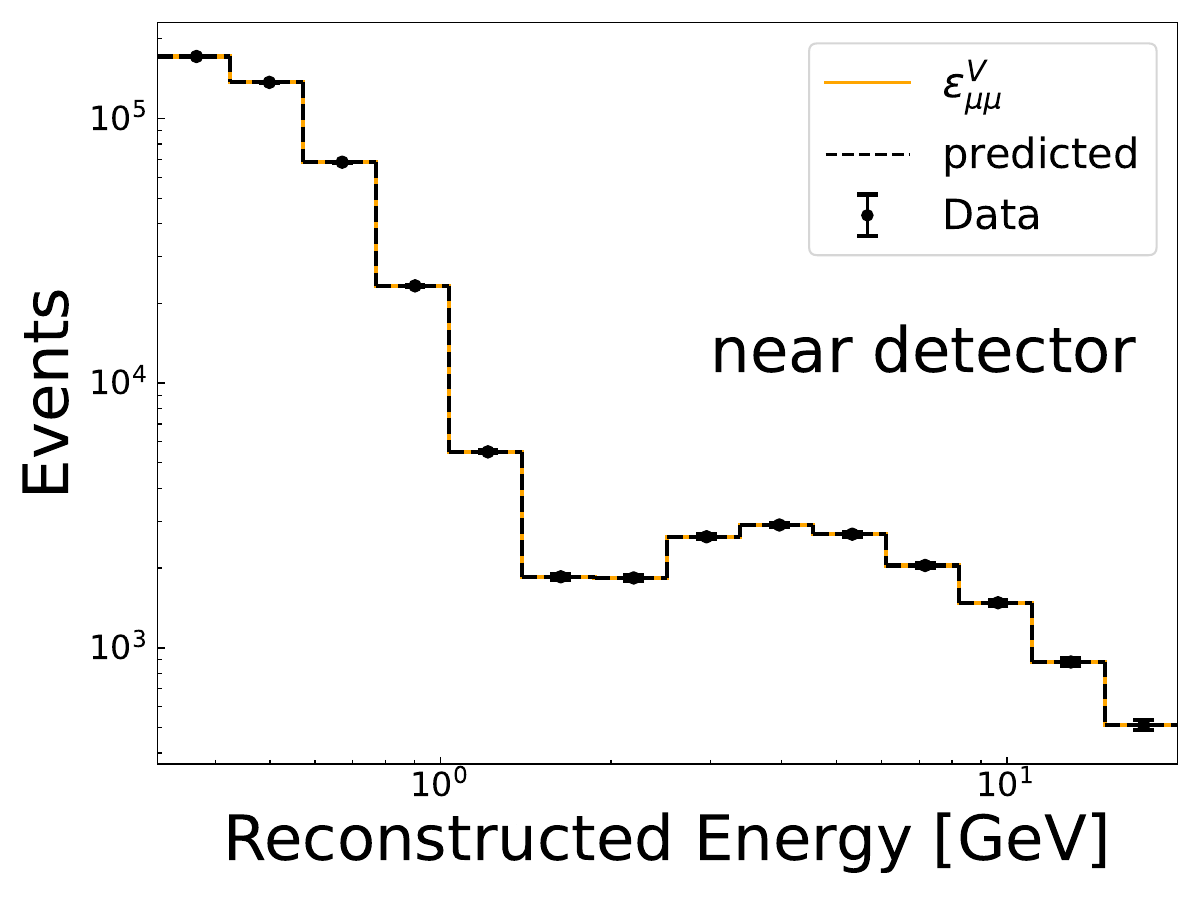}
        \end{adjustbox}
        \begin{adjustbox}{valign=c} 
        \includegraphics[scale=0.33]{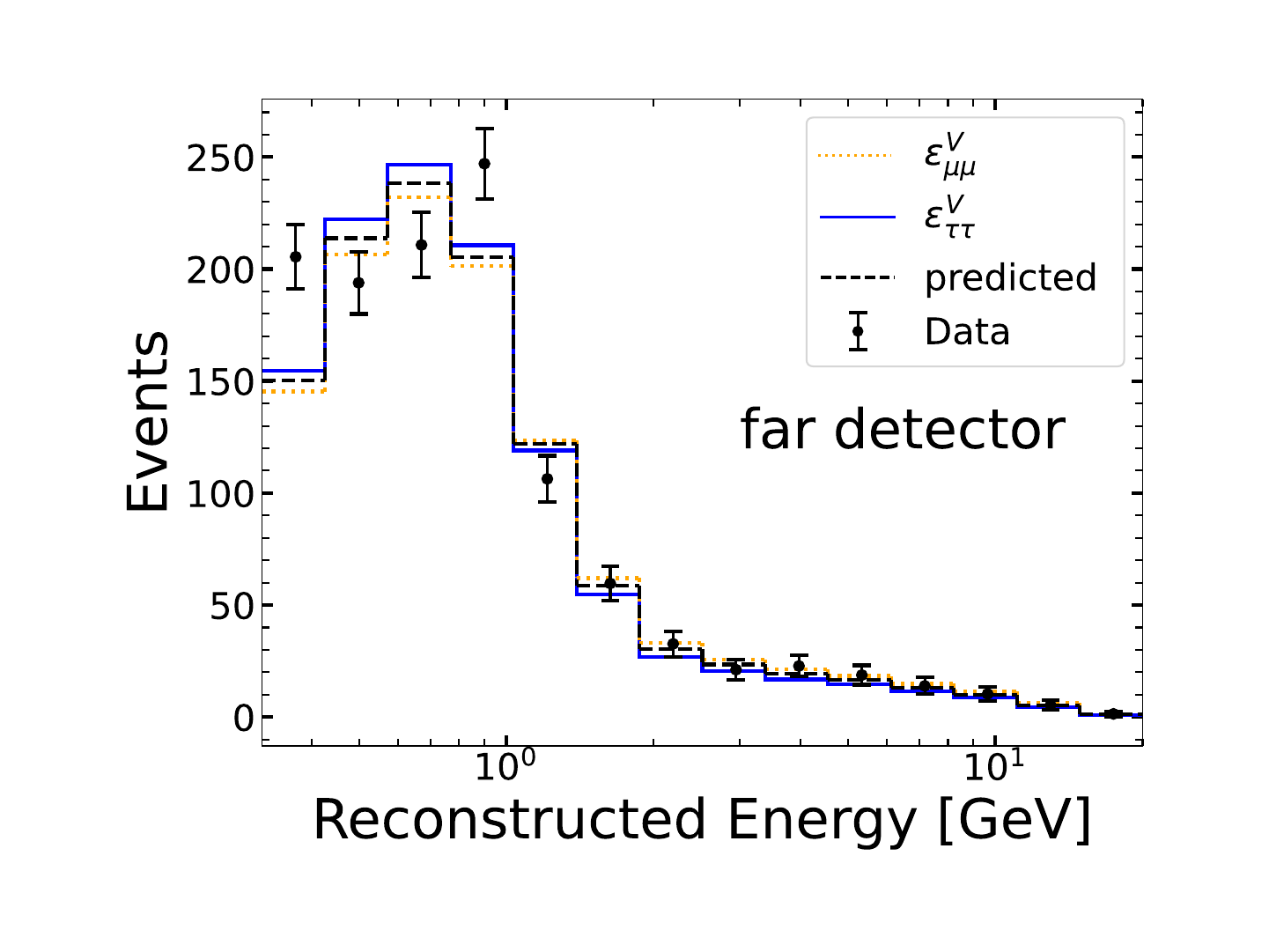}
        \end{adjustbox}
    \caption{
    Neutral current event spectra at NOvA's near and far detectors for the standard prediction (black), and for two benchmark NSI parameters $\vareps_{\mu\mu}^V=0.5$ (orange) and $\vareps_{\tau\tau}^V = 0.7$ (blue).  
    We also show NOvA's data \cite{NOvA:2024imi} including statistical and systematic uncertainties. 
    Note that the near detector spectrum is a perfect fit to data due to our assumption of bin-to-bin uncertainties fully correlated between near and far detectors.
    }
    \label{fig:spectra}
\end{figure}

To understand the experimental impact of NSIs, Fig.~\ref{fig:spectra} shows the reconstructed energy spectra at NOvA for both near and far detectors (left and right panels, respectively) to two non-zero NSI benchmark values, $\varepsilon^V_{\mu\mu} = 0.5$ and $\varepsilon^V_{\tau\tau} = 0.7$. 
Note that, since there are no tau neutrinos in the near detector, to first approximation, we do not show the histogram for $\varepsilon^V_{\tau\tau}$ in the near detector panel. 
Due to the way we structured our fit, allowing every reconstructed energy bin to be fitted freely, the near detector spectrum perfectly matches the data. 
Nevertheless, the impact of NSIs is propagated to the far detector because we have assumed the bin-to-bin uncertainties to be fully correlated between the near and far detectors. 
As shown, there is a non-trivial change in the neutral current reconstructed energy spectrum in the far detector. 
For $\varepsilon_{\mu\mu}^V$, the enhancement of ND events translates into a lower rate of events in the FD, while a positive $\varepsilon_{\tau\tau}^V$ raises the FD events due to the large presence of tau neutrinos from $\nu_\mu\to\nu_\tau$ oscillations.

\begin{figure}
    \centering
    \includegraphics[scale=0.35]{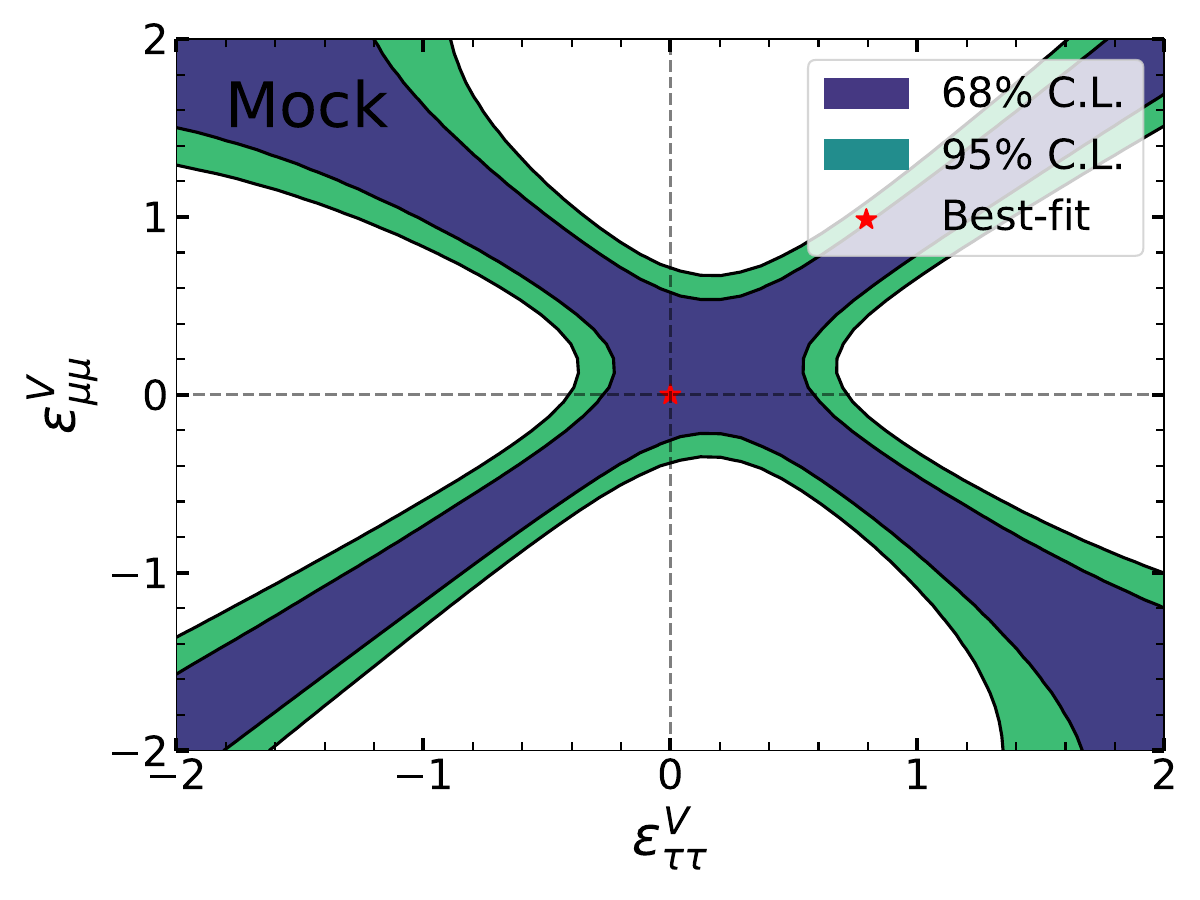}
\includegraphics[scale=0.35]{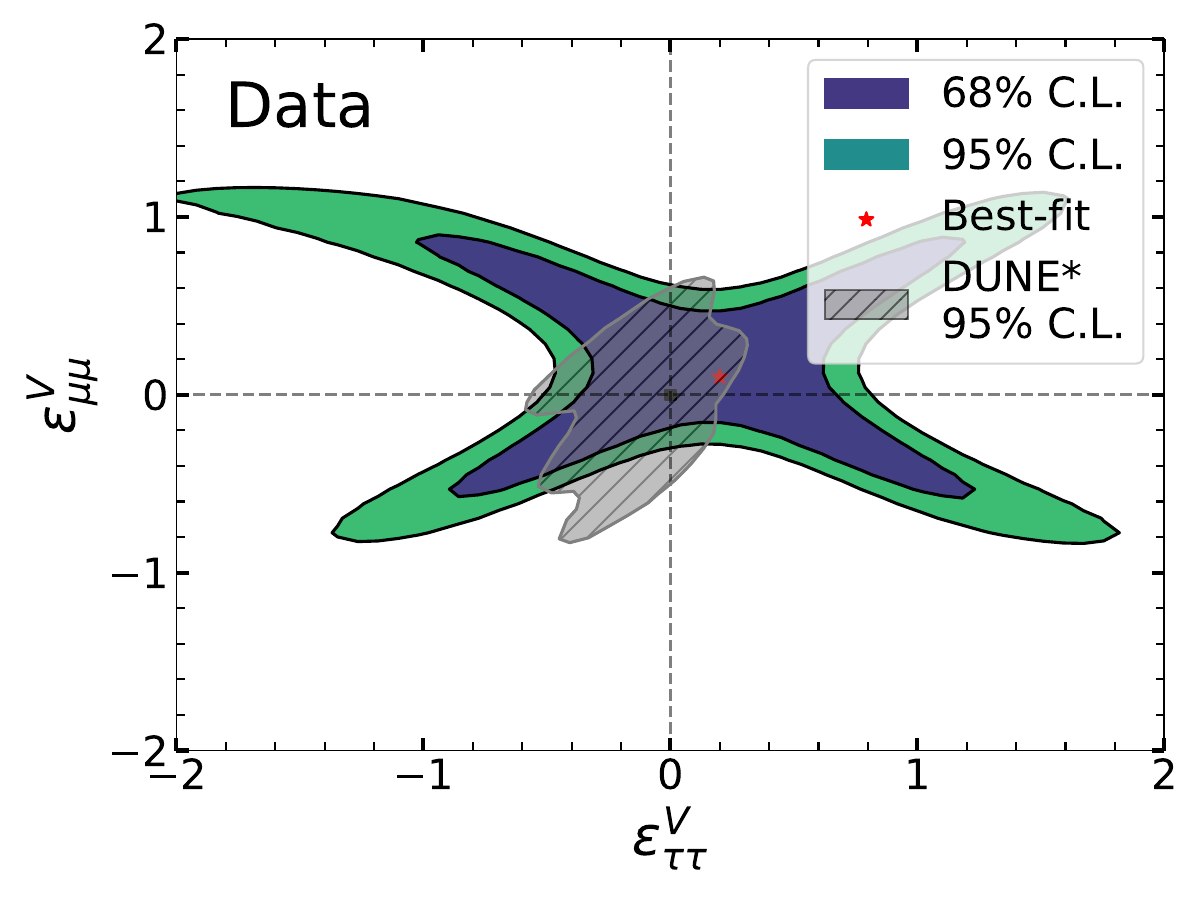}
    \includegraphics[scale=0.35]{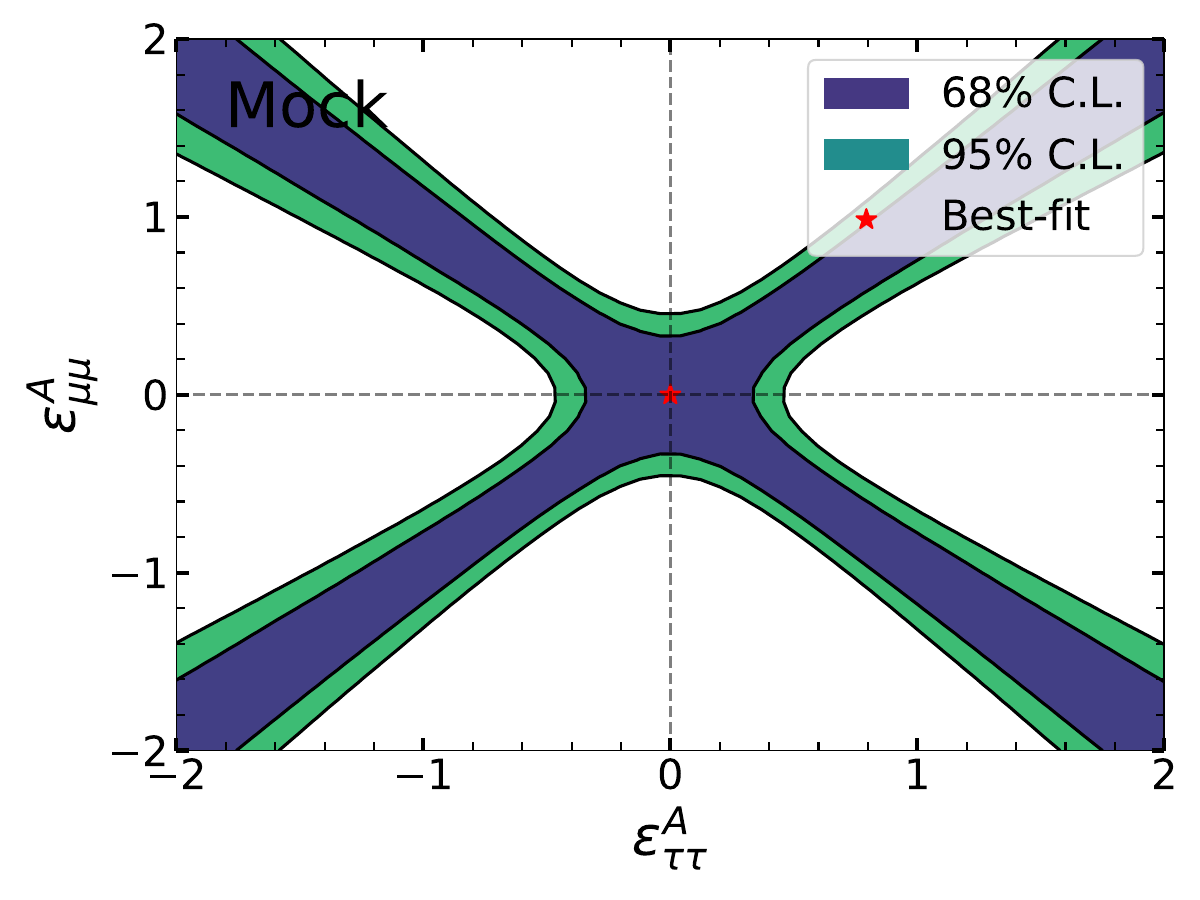}
\includegraphics[scale=0.35]{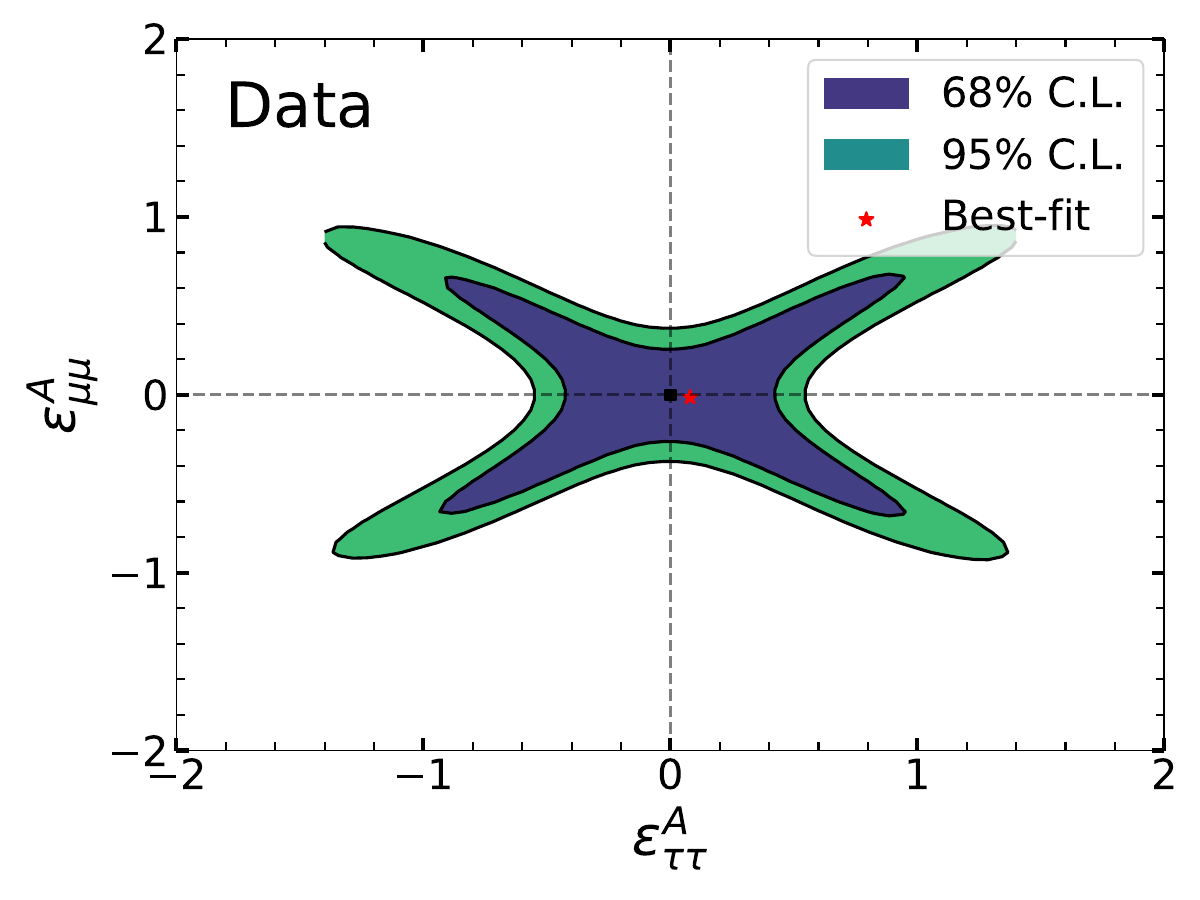}

    \caption{Expected and actual constraints on  $\vareps_{\mu\mu}^V-\vareps_{\tau\tau}^V$ and  $\vareps_{\mu\mu}^A-\vareps_{\tau\tau}^A$ assuming the other NSI parameters to be zero and $13.6\times 10^{20}$ POT for NOvA.  
    For the plots on the left we assume a measurement of the SM (left plots), for the plots on the right we analyzed   existing NOvA data from Ref.~\cite{NOvA:2024imi}. For all plots  we fix the true values of the oscillation parameters on the values from \cite{NOvA:2024imi}.
    The best fit values are $\vareps^V_{\mu\mu}=0.16,~\vareps^V_{\tau\tau}=0.22$ and $\vareps^A_{\mu\mu}=-0.02,~\vareps^A_{\tau\tau}=0.09$. 
    We took the DUNE sensitivity to vectorial NSIs from Ref.~\cite{Chatterjee:2021wac}.
    }
    \label{fig:mm-tt-constraint}
\end{figure}

As mentioned, the constraints presented here are derived by assuming only one non-zero NSI parameter at a time. 
To appreciate the impact of the degeneracies discussed in the previous section, we show in Fig.~\ref{fig:mm-tt-constraint} the experimental sensitivity to both mock (left) and actual (right) NOvA data in the $\varepsilon^V_{\mu\mu} \times \varepsilon^V_{\tau\tau}$ plane (upper panels) as well as the $\varepsilon^A_{\mu\mu} \times \varepsilon^A_{\tau\tau}$ plane (lower panels). We have kept the true values of the oscillation parameters fixed to the values from \cite{NOvA:2024imi}.
First, we observe that the degeneracies described in Eqs.~(\ref{eq:degeneracies}) and (\ref{eq:axial-degeneracies}) are clearly visible.
For the vectorial case, we also show in the actual data plot the estimated sensitivity of the DUNE experiment to $\varepsilon^V_{\mu\mu}$ and $\varepsilon^V_{\tau\tau}$ for an exposure of $336$ kton-MW-year~\cite{Chatterjee:2021wac}. 
As shown, the NOvA neutral current analysis covers a different region of parameter space compared to the future DUNE sensitivity. 
This highlights the inherent synergy between these two approaches, demonstrating the promising potential of neutral current measurements compared to the more commonly used charged current oscillation analyses in probing NSIs. Furthermore, we anticipate even better sensitivity with the future DUNE neutral current measurements \cite{Abbaslu:2023vqk}.~\footnote{A full analysis of the DUNE neutral current measurement for probing NSIs, accounting for the impact of DUNE-PRISM, is left to future work.}
For  axial NSIs, we also observe a degeneracy that is approximately symmetric around the Standard Model expectation, $\varepsilon^A_{\mu\mu} = \varepsilon^A_{\tau\tau} = 0$. 
This feature is visible in both the mock and actual data analyses. 

Overall, we can see  that our approach has significant potential for improving NSI parameter constraints for both vector and axial cases. 
Notably, for axial NSIs, the sensitivity of long-baseline neutral current experiments appears to be highly complementary to that of coherent elastic neutrino-nucleus scattering experiments, offering a comprehensive strategy for probing these interactions.
For completeness, we present the constraints on vector and axial NSIs derived here using NOvA neutral current data in Table~\ref{tab:results}.

Finally, we mention that if NC NSI is realized in nature, adding the   NC detection channel to global fits when extracting the oscillation parameters will be an interesting possibility, as NC events are considered background to the CC channel, and can add valuable information as well as aid in breaking degeneracies between multiple NSI parameters. 

\begin{table}[tb]
\centering
\begin{tabular}{c|c}
vector NSI& constraint at 90\%C.L.\\\hline
$\vareps_{ee}^V$&$[-6.4,1.7]$\\
$\vareps_{\mu\mu}^V$&$[-0.16, 0.47]$\\
$\vareps_{\tau\tau}^V$&$[-0.35 ,0.66]$\\
$\vareps_{e\mu}^V$& $[-0.23, 0.39]$\\
$\vareps_{e\tau}^V$&$[-0.36, 0.49]$\\
$\vareps_{\mu\tau}^V$&$[-0.52, 0.52]$\\
\end{tabular}\hspace{1cm}
\begin{tabular}{c|c}
axial-vector NSI& constraint at 90\%C.L.\\\hline
$\vareps_{ee}^A$&$[-1.6,1.6]$\\
$\vareps_{\mu\mu}^A$&$[-0.25,0.25]$\\
$\vareps_{\tau\tau}^A$&$[-0.41, 0.41]$\\
$\vareps_{e\mu}^A$&$[-0.26,0.26]$\\
$\vareps_{e\tau}^A$&$[-0.40, 0.40]$\\
$\vareps_{\mu\tau}^A$&$[-0.49, 0.49]$
\end{tabular}
\caption{Constraints on vector and axial-vector NSI parameters using existing NOvA neutral current data with 13.6$\times 10^{20}$ POT \cite{NOvA:2024imi} at 90$\%$ C.L. ($\Delta \chi^2=2.71$). We assumed $\epsilon^{V(A)}_{\alpha\beta}=\epsilon^{pV(A)}_{\alpha\beta}=\epsilon^{nV(A)}_{\alpha\beta}$.
We consider  one non-zero NSI parameter at a time and the off-diagonal NSI parameters to be real as there is little sensitivity to their phase. 
}
\label{tab:results}
\end{table}

\section{Conclusions}
\label{sec:conclusions}

In this paper, we have analyzed for the first time the impact of neutral current non-standard interactions (NSIs) in neutral current events at long-baseline neutrino experiments. 
The presence of such NSIs in the cross section introduces a dependence of neutral current events on neutrino flavor, making oscillation physics relevant in the neutral current sample. 
This enables the use of near-to-far detector correlations to mitigate the large systematic uncertainties associated with the neutrino flux and interaction cross section.

As a concrete demonstration of the capability of neutral current events to constrain NSIs, we have analyzed NOvA’s present neutral current data. 
Our results show that NOvA can improve existing constraints on both vector and axial-vector NSIs. 
In particular, NOvA’s neutral current data significantly tighten constraints on $\varepsilon_{\mu\mu}^{A}$, $\varepsilon_{\tau\tau}^{A}$, and $|\varepsilon_{e\mu}^A|$, compared to current global fit results, and together with COHERENT and DRESDEN-II, it helps to exclude the LMA-Dark region for $\varepsilon_{\mu\mu}^V$ and $\varepsilon_{\tau\tau}^V$.
It is important to note that short baseline experiments probing neutral current interactions, like COHERENT and DRESDEN-II, are sensitive to NSIs with electron and muon neutrinos.
On the other hand, while our proposed long-baseline neutral current analysis can constrain NSIs with muon and tau neutrinos.
Combining these two strategies allows for probing new interactions with all neutrino flavors, specially for axial-vector couplings.
We have also highlighted a potential synergy between neutral and charged current samples in long-baseline experiments, which can help resolve degeneracies in vector NSIs,  distinguish between axial \(\varepsilon_u^A\) and \(\varepsilon_d^A\) NSIs and generating the first bounds on the isospin conserving scenario \(\varepsilon_d^A=\varepsilon_u^A\).

\section*{Acknowledgments}
We thank Alex Himmel for invaluable discussions on neutrino energy reconstruction at NOvA and Salvador Urrea for his constructive comments.
JG acknowledges support by the U.S. Department of Energy Office of Science under award number  DE-SC0025448. 
JPP is supported by grant  PID2022-\allowbreak 126224NB-\allowbreak C21 and  "Unit of Excellence Maria de Maeztu 2020-2023'' award to the ICC-UB CEX2019-000918-M  funded by MCIN/AEI/\allowbreak 10.13039/\allowbreak 501100011033, 
also supported by the European Union's through the
Horizon 2020 research and innovation program (Marie
Sk{\l}odowska-Curie grant agreement 860881-HIDDeN) and the Horizon
Europe research and innovation programme (Marie Sk{\l}odowska-Curie
Staff Exchange grant agreement 101086085-ASYMMETRY). 
It also receives support  from grant 2021-SGR-249 (Generalitat de Catalunya).
Fermilab is operated by the Fermi Research Alliance, LLC under contract No. DE-AC02-07CH11359 with the United States Department of Energy.

\appendix
\section{SM prediction of NC cross section}
\label{sec:smnc}
NOvA's near and far detectors   use $\mathrm{CH}_2$ as detector material.
We use \texttt{NuWro} \cite{Juszczak:2005zs,Golan:2012wx,Golan:2012rfa} to calculate the SM expectation for the NC scattering process 
$$\nu_\mu \, \mathrm{CH}_2 \to \nu X~.$$ 
The NC interaction has contributions from quasi-elastic scattering (QE), resonant scattering (RES) and deep inelastic scattering (DIS).
We use as input parameters to NuWro 
\begin{eqnarray}
    &&M_A^{\mathrm{QE}}=0.99\,\mathrm{GeV}\nonumber\\
    &&M_A^{\pi}=0.94\,\mathrm{GeV}
\end{eqnarray}

\begin{figure}
    \centering
\includegraphics[scale=0.4]{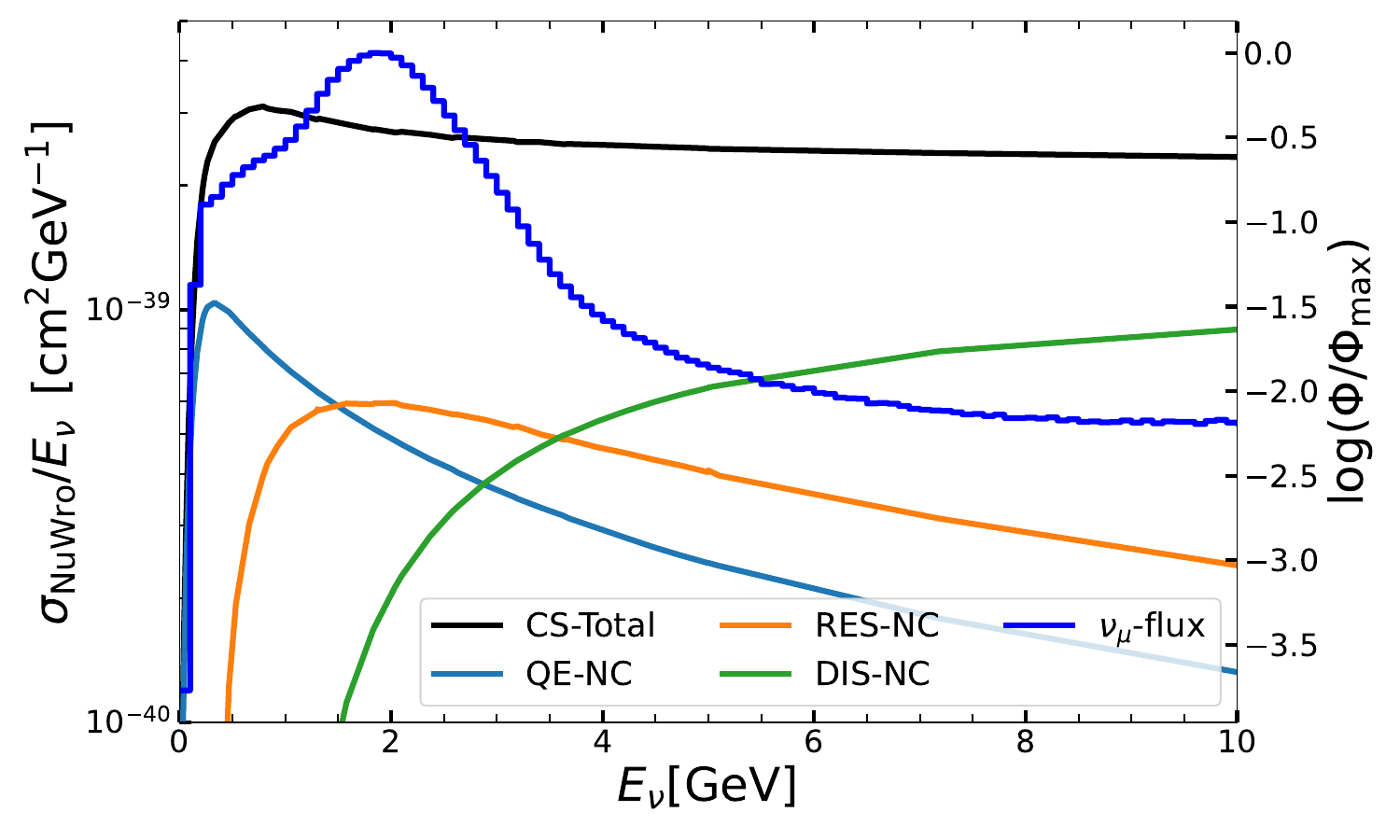}
    \caption{The neutral current cross section in the SM calculated with \texttt{NuWro} as a function of the neutrino energy. We show the individual contributions to the cross section and the NOvA  $\nu_\mu$ flux $\phi$ for comparison (see App.~\ref{sec:smnc} for details). 
    }
    \label{fig:xsecs}
\end{figure}

In Fig.~\ref{fig:xsecs} we show these cross section components as well as the total NC cross section and the neutrino flux at NOvA for reference.

\section{Migration matrix}

\begin{figure}
    \centering
\includegraphics[scale=0.4]{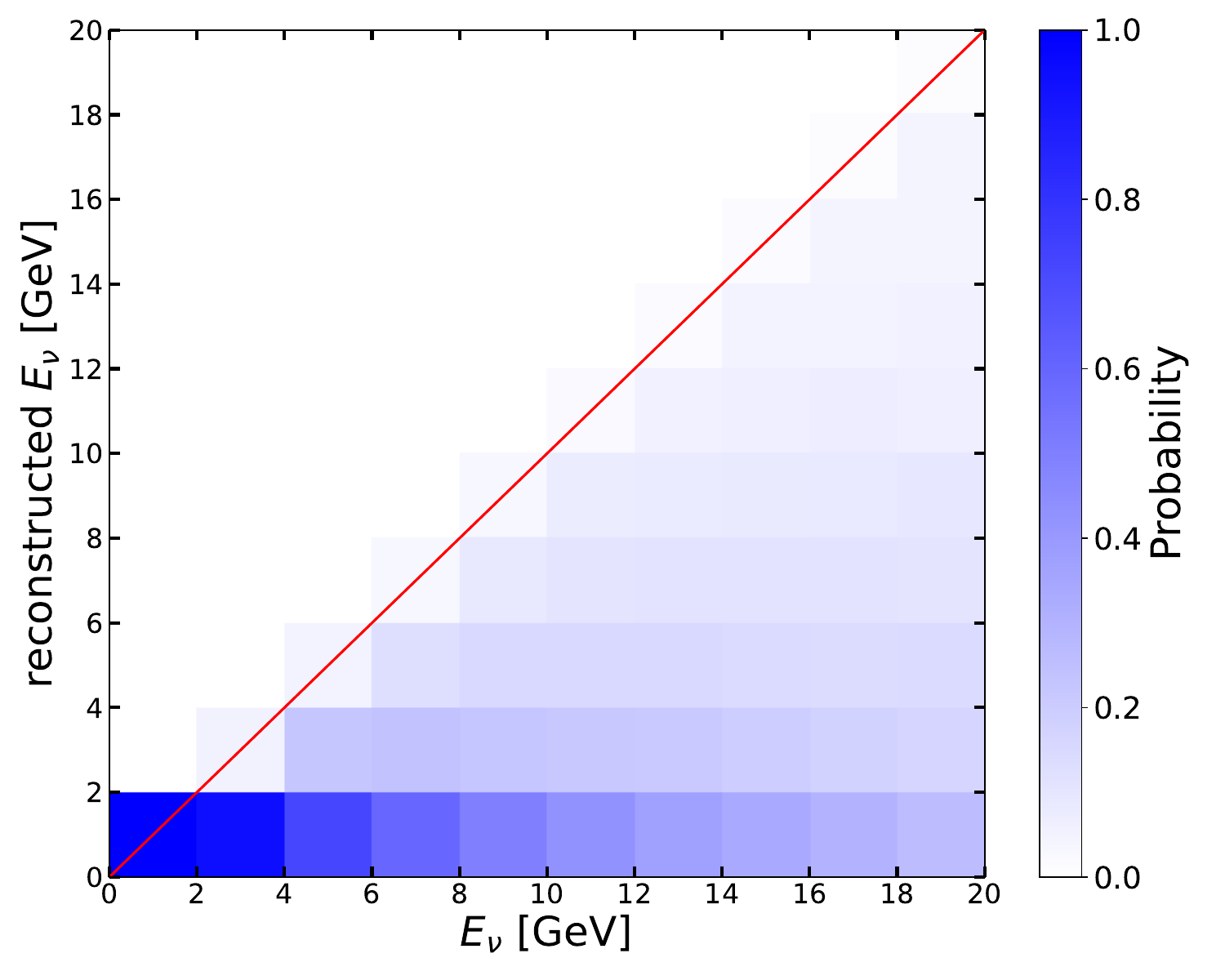}
    \caption{The migration matrix between true and reconstructed neutrino energies which we use for our analys  is of the NOvA data \cite{NOvA:2024imi}.  }
    \label{fig:migration}
\end{figure}

\label{sec:migration}

For the conversion between true and reconstructed neutrino energies, we follow the approach from Ref.~\cite{Rodrigues:2016xjj}. 
We calculate the reconstructed energy at an event-by-event basis.
Figure~\ref{fig:migration} shows the migration matrix between the real neutrino energy and the reconstructed energy.

\bibliographystyle{JHEP}
\bibliography{main}

\providecommand{\href}[2]{#2}\begingroup\raggedright\begin{thebibliography}{10}

\bibitem{Heeck:2011wj}
J.~Heeck and W.~Rodejohann, \emph{{Gauged $L_\mu - L_\tau$ Symmetry at the
  Electroweak Scale}},
  \href{https://doi.org/10.1103/PhysRevD.84.075007}{\emph{Phys. Rev. D}
  {\bfseries 84} (2011) 075007}
  [\href{https://arxiv.org/abs/1107.5238}{{\ttfamily 1107.5238}}].

\bibitem{Farzan:2015doa}
Y.~Farzan, \emph{{A model for large non-standard interactions of neutrinos
  leading to the LMA-Dark solution}},
  \href{https://doi.org/10.1016/j.physletb.2015.07.015}{\emph{Phys. Lett. B}
  {\bfseries 748} (2015) 311}
  [\href{https://arxiv.org/abs/1505.06906}{{\ttfamily 1505.06906}}].

\bibitem{Farzan:2015hkd}
Y.~Farzan and I.M.~Shoemaker, \emph{{Lepton Flavor Violating Non-Standard
  Interactions via Light Mediators}},
  \href{https://doi.org/10.1007/JHEP07(2016)033}{\emph{JHEP} {\bfseries 07}
  (2016) 033} [\href{https://arxiv.org/abs/1512.09147}{{\ttfamily
  1512.09147}}].

\bibitem{Farzan:2016wym}
Y.~Farzan and J.~Heeck, \emph{{Neutrinophilic nonstandard interactions}},
  \href{https://doi.org/10.1103/PhysRevD.94.053010}{\emph{Phys. Rev. D}
  {\bfseries 94} (2016) 053010}
  [\href{https://arxiv.org/abs/1607.07616}{{\ttfamily 1607.07616}}].

\bibitem{Forero:2016ghr}
D.V.~Forero and W.-C.~Huang, \emph{{Sizable NSI from the SU(2)$_{L}$ scalar
  doublet-singlet mixing and the implications in DUNE}},
  \href{https://doi.org/10.1007/JHEP03(2017)018}{\emph{JHEP} {\bfseries 03}
  (2017) 018} [\href{https://arxiv.org/abs/1608.04719}{{\ttfamily
  1608.04719}}].

\bibitem{Babu:2017olk}
K.S.~Babu, A.~Friedland, P.A.N.~Machado and I.~Mocioiu, \emph{{Flavor Gauge
  Models Below the Fermi Scale}},
  \href{https://doi.org/10.1007/JHEP12(2017)096}{\emph{JHEP} {\bfseries 12}
  (2017) 096} [\href{https://arxiv.org/abs/1705.01822}{{\ttfamily
  1705.01822}}].

\bibitem{Heeck:2018nzc}
J.~Heeck, M.~Lindner, W.~Rodejohann and S.~Vogl, \emph{{Non-Standard Neutrino
  Interactions and Neutral Gauge Bosons}},
  \href{https://doi.org/10.21468/SciPostPhys.6.3.038}{\emph{SciPost Phys.}
  {\bfseries 6} (2019) 038} [\href{https://arxiv.org/abs/1812.04067}{{\ttfamily
  1812.04067}}].

\bibitem{Denton:2018dqq}
P.B.~Denton, Y.~Farzan and I.M.~Shoemaker, \emph{{Activating the fourth
  neutrino of the 3+1 scheme}},
  \href{https://doi.org/10.1103/PhysRevD.99.035003}{\emph{Phys. Rev. D}
  {\bfseries 99} (2019) 035003}
  [\href{https://arxiv.org/abs/1811.01310}{{\ttfamily 1811.01310}}].

\bibitem{Dey:2018yht}
U.K.~Dey, N.~Nath and S.~Sadhukhan, \emph{{Non-Standard Neutrino Interactions
  in a Modified $\nu$2HDM}},
  \href{https://doi.org/10.1103/PhysRevD.98.055004}{\emph{Phys. Rev. D}
  {\bfseries 98} (2018) 055004}
  [\href{https://arxiv.org/abs/1804.05808}{{\ttfamily 1804.05808}}].

\bibitem{Babu:2019mfe}
K.S.~Babu, P.S.B.~Dev, S.~Jana and A.~Thapa, \emph{{Non-Standard Interactions
  in Radiative Neutrino Mass Models}},
  \href{https://doi.org/10.1007/JHEP03(2020)006}{\emph{JHEP} {\bfseries 03}
  (2020) 006} [\href{https://arxiv.org/abs/1907.09498}{{\ttfamily
  1907.09498}}].

\bibitem{Babu:2020nna}
K.S.~Babu, D.~Gon\c{c}alves, S.~Jana and P.A.N.~Machado, \emph{{Neutrino
  Non-Standard Interactions: Complementarity Between LHC and Oscillation
  Experiments}},
  \href{https://doi.org/10.1016/j.physletb.2021.136131}{\emph{Phys. Lett. B}
  {\bfseries 815} (2021) 136131}
  [\href{https://arxiv.org/abs/2003.03383}{{\ttfamily 2003.03383}}].

\bibitem{Babu:2021cxe}
K.S.~Babu, V.~Brdar, A.~de~Gouv\^ea and P.A.N.~Machado, \emph{{Energy-dependent
  neutrino mixing parameters at oscillation experiments}},
  \href{https://doi.org/10.1103/PhysRevD.105.115014}{\emph{Phys. Rev. D}
  {\bfseries 105} (2022) 115014}
  [\href{https://arxiv.org/abs/2108.11961}{{\ttfamily 2108.11961}}].

\bibitem{Falkowski:2021bkq}
A.~Falkowski, M.~Gonz\'alez-Alonso, J.~Kopp, Y.~Soreq and Z.~Tabrizi,
  \emph{{EFT at FASER\ensuremath{\nu}}},
  \href{https://doi.org/10.1007/JHEP10(2021)086}{\emph{JHEP} {\bfseries 10}
  (2021) 086} [\href{https://arxiv.org/abs/2105.12136}{{\ttfamily
  2105.12136}}].

\bibitem{Coloma:2023ixt}
P.~Coloma, M.C.~Gonzalez-Garcia, M.~Maltoni, J.a.P.~Pinheiro and S.~Urrea,
  \emph{{Global constraints on non-standard neutrino interactions with quarks
  and electrons}}, \href{https://doi.org/10.1007/JHEP08(2023)032}{\emph{JHEP}
  {\bfseries 08} (2023) 032}
  [\href{https://arxiv.org/abs/2305.07698}{{\ttfamily 2305.07698}}].

\bibitem{NOvA:2024lti}
{\scshape NOvA} collaboration, \emph{{Search for CP-violating Neutrino
  Non-Standard Interactions with the NOvA Experiment}},
  \href{https://arxiv.org/abs/2403.07266}{{\ttfamily 2403.07266}}.

\bibitem{MINOS:2013hmj}
{\scshape MINOS} collaboration, \emph{{Search for Flavor-Changing Non-Standard
  Neutrino Interactions by MINOS}},
  \href{https://doi.org/10.1103/PhysRevD.88.072011}{\emph{Phys. Rev. D}
  {\bfseries 88} (2013) 072011}
  [\href{https://arxiv.org/abs/1303.5314}{{\ttfamily 1303.5314}}].

\bibitem{IceCube:2017zcu}
{\scshape IceCube} collaboration, \emph{{Search for Nonstandard Neutrino
  Interactions with IceCube DeepCore}},
  \href{https://doi.org/10.1103/PhysRevD.97.072009}{\emph{Phys. Rev. D}
  {\bfseries 97} (2018) 072009}
  [\href{https://arxiv.org/abs/1709.07079}{{\ttfamily 1709.07079}}].

\bibitem{IceCubeCollaboration:2021euf}
{\scshape (IceCube Collaboration)*, IceCube} collaboration, \emph{{All-flavor
  constraints on nonstandard neutrino interactions and generalized matter
  potential with three years of IceCube DeepCore data}},
  \href{https://doi.org/10.1103/PhysRevD.104.072006}{\emph{Phys. Rev. D}
  {\bfseries 104} (2021) 072006}
  [\href{https://arxiv.org/abs/2106.07755}{{\ttfamily 2106.07755}}].

\bibitem{IceCube:2022ubv}
{\scshape IceCube} collaboration, \emph{{Strong Constraints on Neutrino
  Nonstandard Interactions from TeV-Scale $\nu_u$ Disappearance at IceCube}},
  \href{https://doi.org/10.1103/PhysRevLett.129.011804}{\emph{Phys. Rev. Lett.}
  {\bfseries 129} (2022) 011804}
  [\href{https://arxiv.org/abs/2201.03566}{{\ttfamily 2201.03566}}].

\bibitem{Super-Kamiokande:2011dam}
{\scshape Super-Kamiokande} collaboration, \emph{{Study of Non-Standard
  Neutrino Interactions with Atmospheric Neutrino Data in Super-Kamiokande I
  and II}}, \href{https://doi.org/10.1103/PhysRevD.84.113008}{\emph{Phys. Rev.
  D} {\bfseries 84} (2011) 113008}
  [\href{https://arxiv.org/abs/1109.1889}{{\ttfamily 1109.1889}}].

\bibitem{Super-Kamiokande:2022lyl}
{\scshape Super-Kamiokande} collaboration, \emph{{Testing Non-Standard
  Interactions Between Solar Neutrinos and Quarks with Super-Kamiokande}},
  \href{https://arxiv.org/abs/2203.11772}{{\ttfamily 2203.11772}}.

\bibitem{Proceedings:2019qno}
\emph{{Neutrino Non-Standard Interactions: A Status Report}}, vol.~2, 2019.
\newblock 10.21468/SciPostPhysProc.2.001.

\bibitem{Grifols:2003gy}
J.A.~Grifols and E.~Masso, \emph{{Neutrino oscillations in the sun probe long
  range leptonic forces}},
  \href{https://doi.org/10.1016/j.physletb.2003.10.078}{\emph{Phys. Lett. B}
  {\bfseries 579} (2004) 123}
  [\href{https://arxiv.org/abs/hep-ph/0311141}{{\ttfamily hep-ph/0311141}}].

\bibitem{Joshipura:2003jh}
A.S.~Joshipura and S.~Mohanty, \emph{{Constraints on flavor dependent long
  range forces from atmospheric neutrino observations at super-Kamiokande}},
  \href{https://doi.org/10.1016/j.physletb.2004.01.057}{\emph{Phys. Lett. B}
  {\bfseries 584} (2004) 103}
  [\href{https://arxiv.org/abs/hep-ph/0310210}{{\ttfamily hep-ph/0310210}}].

\bibitem{Davoudiasl:2011sz}
H.~Davoudiasl, H.-S.~Lee and W.J.~Marciano, \emph{{Long-Range Lepton Flavor
  Interactions and Neutrino Oscillations}},
  \href{https://doi.org/10.1103/PhysRevD.84.013009}{\emph{Phys. Rev. D}
  {\bfseries 84} (2011) 013009}
  [\href{https://arxiv.org/abs/1102.5352}{{\ttfamily 1102.5352}}].

\bibitem{Wise:2018rnb}
M.B.~Wise and Y.~Zhang, \emph{{Lepton Flavorful Fifth Force and Depth-dependent
  Neutrino Matter Interactions}},
  \href{https://doi.org/10.1007/JHEP06(2018)053}{\emph{JHEP} {\bfseries 06}
  (2018) 053} [\href{https://arxiv.org/abs/1803.00591}{{\ttfamily
  1803.00591}}].

\bibitem{Smirnov:2019cae}
A.Y.~Smirnov and X.-J.~Xu, \emph{{Wolfenstein potentials for neutrinos induced
  by ultra-light mediators}},
  \href{https://doi.org/10.1007/JHEP12(2019)046}{\emph{JHEP} {\bfseries 12}
  (2019) 046} [\href{https://arxiv.org/abs/1909.07505}{{\ttfamily
  1909.07505}}].

\bibitem{Babu:2019iml}
K.S.~Babu, G.~Chauhan and P.S.~Bhupal~Dev, \emph{{Neutrino nonstandard
  interactions via light scalars in the Earth, Sun, supernovae, and the early
  Universe}}, \href{https://doi.org/10.1103/PhysRevD.101.095029}{\emph{Phys.
  Rev. D} {\bfseries 101} (2020) 095029}
  [\href{https://arxiv.org/abs/1912.13488}{{\ttfamily 1912.13488}}].

\bibitem{Coloma:2020gfv}
P.~Coloma, M.C.~Gonzalez-Garcia and M.~Maltoni, \emph{{Neutrino oscillation
  constraints on U(1)' models: from non-standard interactions to long-range
  forces}}, \href{https://doi.org/10.1007/JHEP01(2021)114}{\emph{JHEP}
  {\bfseries 01} (2021) 114}
  [\href{https://arxiv.org/abs/2009.14220}{{\ttfamily 2009.14220}}].

\bibitem{Chauhan:2024qew}
G.~Chauhan and X.-J.~Xu, \emph{{Impact of the cosmic neutrino background on
  long-range force searches}},
  \href{https://doi.org/10.1007/JHEP07(2024)255}{\emph{JHEP} {\bfseries 07}
  (2024) 255} [\href{https://arxiv.org/abs/2403.09783}{{\ttfamily
  2403.09783}}].

\bibitem{Dutta:2020che}
B.~Dutta, R.F.~Lang, S.~Liao, S.~Sinha, L.~Strigari and A.~Thompson, \emph{{A
  global analysis strategy to resolve neutrino NSI degeneracies with scattering
  and oscillation data}},
  \href{https://doi.org/10.1007/JHEP09(2020)106}{\emph{JHEP} {\bfseries 09}
  (2020) 106} [\href{https://arxiv.org/abs/2002.03066}{{\ttfamily
  2002.03066}}].

\bibitem{Coloma:2022umy}
P.~Coloma, M.C.~Gonzalez-Garcia, M.~Maltoni, J.a.P.~Pinheiro and S.~Urrea,
  \emph{{Constraining new physics with Borexino Phase-II spectral data}},
  \href{https://doi.org/10.1007/JHEP07(2022)138}{\emph{JHEP} {\bfseries 07}
  (2022) 138} [\href{https://arxiv.org/abs/2204.03011}{{\ttfamily
  2204.03011}}].

\bibitem{COHERENT:2017ipa}
{\scshape COHERENT} collaboration, \emph{{Observation of Coherent Elastic
  Neutrino-Nucleus Scattering}},
  \href{https://doi.org/10.1126/science.aao0990}{\emph{Science} {\bfseries 357}
  (2017) 1123} [\href{https://arxiv.org/abs/1708.01294}{{\ttfamily
  1708.01294}}].

\bibitem{COHERENT:2020iec}
{\scshape COHERENT} collaboration, \emph{{First Measurement of Coherent Elastic
  Neutrino-Nucleus Scattering on Argon}},
  \href{https://doi.org/10.1103/PhysRevLett.126.012002}{\emph{Phys. Rev. Lett.}
  {\bfseries 126} (2021) 012002}
  [\href{https://arxiv.org/abs/2003.10630}{{\ttfamily 2003.10630}}].

\bibitem{COHERENT:2021xmm}
{\scshape COHERENT} collaboration, \emph{{Measurement of the Coherent Elastic
  Neutrino-Nucleus Scattering Cross Section on CsI by COHERENT}},
  \href{https://doi.org/10.1103/PhysRevLett.129.081801}{\emph{Phys. Rev. Lett.}
  {\bfseries 129} (2022) 081801}
  [\href{https://arxiv.org/abs/2110.07730}{{\ttfamily 2110.07730}}].

\bibitem{Adamski:2024yqt}
S.~Adamski et~al., \emph{{First detection of coherent elastic neutrino-nucleus
  scattering on germanium}},
  \href{https://arxiv.org/abs/2406.13806}{{\ttfamily 2406.13806}}.

\bibitem{Colaresi:2022obx}
J.~Colaresi, J.I.~Collar, T.W.~Hossbach, C.M.~Lewis and K.M.~Yocum,
  \emph{{Measurement of Coherent Elastic Neutrino-Nucleus Scattering from
  Reactor Antineutrinos}},
  \href{https://doi.org/10.1103/PhysRevLett.129.211802}{\emph{Phys. Rev. Lett.}
  {\bfseries 129} (2022) 211802}
  [\href{https://arxiv.org/abs/2202.09672}{{\ttfamily 2202.09672}}].

\bibitem{MINOS:2008ccf}
{\scshape MINOS} collaboration, \emph{{Search for active neutrino disappearance
  using neutral-current interactions in the MINOS long-baseline experiment}},
  \href{https://doi.org/10.1103/PhysRevLett.101.221804}{\emph{Phys. Rev. Lett.}
  {\bfseries 101} (2008) 221804}
  [\href{https://arxiv.org/abs/0807.2424}{{\ttfamily 0807.2424}}].

\bibitem{NOvA:2021smv}
{\scshape NOvA} collaboration, \emph{{Search for Active-Sterile Antineutrino
  Mixing Using Neutral-Current Interactions with the NOvA Experiment}},
  \href{https://doi.org/10.1103/PhysRevLett.127.201801}{\emph{Phys. Rev. Lett.}
  {\bfseries 127} (2021) 201801}
  [\href{https://arxiv.org/abs/2106.04673}{{\ttfamily 2106.04673}}].

\bibitem{NOvA:2024imi}
{\scshape NOvA} collaboration, \emph{{Dual-Baseline Search for
  Active-to-Sterile Neutrino Oscillations in NOvA}},
  \href{https://arxiv.org/abs/2409.04553}{{\ttfamily 2409.04553}}.

\bibitem{Gandhi:2017vzo}
R.~Gandhi, B.~Kayser, S.~Prakash and S.~Roy, \emph{{What measurements of
  neutrino neutral current events can reveal}},
  \href{https://doi.org/10.1007/JHEP11(2017)202}{\emph{JHEP} {\bfseries 11}
  (2017) 202} [\href{https://arxiv.org/abs/1708.01816}{{\ttfamily
  1708.01816}}].

\bibitem{Coloma:2017ptb}
P.~Coloma, D.V.~Forero and S.J.~Parke, \emph{{DUNE Sensitivities to the Mixing
  between Sterile and Tau Neutrinos}},
  \href{https://doi.org/10.1007/JHEP07(2018)079}{\emph{JHEP} {\bfseries 07}
  (2018) 079} [\href{https://arxiv.org/abs/1707.05348}{{\ttfamily
  1707.05348}}].

\bibitem{Dutta:2019hmb}
D.~Dutta and S.~Roy, \emph{{Non-Unitarity at DUNE and T2HK with Charged and
  Neutral Current Measurements}},
  \href{https://doi.org/10.1088/1361-6471/abdc03}{\emph{J. Phys. G} {\bfseries
  48} (2021) 045004} [\href{https://arxiv.org/abs/1901.11298}{{\ttfamily
  1901.11298}}].

\bibitem{Abbaslu:2023vqk}
S.~Abbaslu, M.~Dehpour, Y.~Farzan and S.~Safari, \emph{{Searching for axial
  neutral current non-standard interactions of neutrinos by DUNE-like
  experiments}}, \href{https://doi.org/10.1007/JHEP04(2024)038}{\emph{JHEP}
  {\bfseries 04} (2024) 038}
  [\href{https://arxiv.org/abs/2312.12420}{{\ttfamily 2312.12420}}].

\bibitem{Pontecorvo:1957cp}
B.~Pontecorvo, \emph{{Mesonium and anti-mesonium}}, {\emph{Sov. Phys. JETP}
  {\bfseries 6} (1957) 429}.

\bibitem{Maki:1962mu}
Z.~Maki, M.~Nakagawa and S.~Sakata, \emph{{Remarks on the unified model of
  elementary particles}}, \href{https://doi.org/10.1143/PTP.28.870}{\emph{Prog.
  Theor. Phys.} {\bfseries 28} (1962) 870}.

\bibitem{NOvA:2004blv}
{\scshape NOvA} collaboration, \emph{{NOvA: Proposal to Build a 30 Kiloton
  Off-Axis Detector to Study $\nu_{\mu} \to \nu_e$ Oscillations in the NuMI
  Beamline}},  \href{https://arxiv.org/abs/hep-ex/0503053}{{\ttfamily
  hep-ex/0503053}}.

\bibitem{Amaral:2023tbs}
D.W.P.~Amaral, D.~Cerdeno, A.~Cheek and P.~Foldenauer, \emph{{A direct
  detection view of the neutrino NSI landscape}},
  \href{https://doi.org/10.1007/JHEP07(2023)071}{\emph{JHEP} {\bfseries 07}
  (2023) 071} [\href{https://arxiv.org/abs/2302.12846}{{\ttfamily
  2302.12846}}].

\bibitem{Isaacson:2020wlx}
J.~Isaacson, W.I.~Jay, A.~Lovato, P.A.N.~Machado and N.~Rocco, \emph{{New
  approach to intranuclear cascades with quantum Monte Carlo configurations}},
  \href{https://doi.org/10.1103/PhysRevC.103.015502}{\emph{Phys. Rev. C}
  {\bfseries 103} (2021) 015502}
  [\href{https://arxiv.org/abs/2007.15570}{{\ttfamily 2007.15570}}].

\bibitem{Isaacson:2022cwh}
J.~Isaacson, W.I.~Jay, A.~Lovato, P.A.N.~Machado and N.~Rocco,
  \emph{{Introducing a novel event generator for electron-nucleus and
  neutrino-nucleus scattering}},
  \href{https://doi.org/10.1103/PhysRevD.107.033007}{\emph{Phys. Rev. D}
  {\bfseries 107} (2023) 033007}
  [\href{https://arxiv.org/abs/2205.06378}{{\ttfamily 2205.06378}}].

\bibitem{Juszczak:2005zs}
C.~Juszczak, J.A.~Nowak and J.T.~Sobczyk, \emph{{Simulations from a new
  neutrino event generator}},
  \href{https://doi.org/10.1016/j.nuclphysbps.2006.08.069}{\emph{Nucl. Phys. B
  Proc. Suppl.} {\bfseries 159} (2006) 211}
  [\href{https://arxiv.org/abs/hep-ph/0512365}{{\ttfamily hep-ph/0512365}}].

\bibitem{Golan:2012wx}
T.~Golan, C.~Juszczak and J.T.~Sobczyk, \emph{{Final State Interactions Effects
  in Neutrino-Nucleus Interactions}},
  \href{https://doi.org/10.1103/PhysRevC.86.015505}{\emph{Phys. Rev. C}
  {\bfseries 86} (2012) 015505}
  [\href{https://arxiv.org/abs/1202.4197}{{\ttfamily 1202.4197}}].

\bibitem{Golan:2012rfa}
T.~Golan, J.T.~Sobczyk and J.~Zmuda, \emph{{NuWro: the Wroclaw Monte Carlo
  Generator of Neutrino Interactions}},
  \href{https://doi.org/10.1016/j.nuclphysbps.2012.09.136}{\emph{Nucl. Phys. B
  Proc. Suppl.} {\bfseries 229-232} (2012) 499}.

\bibitem{Kopp:2024yvh}
J.~Kopp, N.~Rocco and Z.~Tabrizi, \emph{{Unleashing the Power of EFT in
  Neutrino-Nucleus Scattering}},
  \href{https://arxiv.org/abs/2401.07902}{{\ttfamily 2401.07902}}.

\bibitem{Ilma:2024lkp}
Ilma, M.~Rafi~Alam, L.~Alvarez-Ruso, M.B.~Galan, I.~Ruiz~Simo and S.K.~Singh,
  \emph{{Neutrino-nucleon elastic scattering in presence of non-standard
  interactions: cross sections and nucleon polarizations}},
  \href{https://arxiv.org/abs/2412.04818}{{\ttfamily 2412.04818}}.

\bibitem{deGouvea:2000pqg}
A.~de~Gouvea, A.~Friedland and H.~Murayama, \emph{{The Dark side of the solar
  neutrino parameter space}},
  \href{https://doi.org/10.1016/S0370-2693(00)00989-8}{\emph{Phys. Lett. B}
  {\bfseries 490} (2000) 125}
  [\href{https://arxiv.org/abs/hep-ph/0002064}{{\ttfamily hep-ph/0002064}}].

\bibitem{Miranda:2004nb}
O.G.~Miranda, M.A.~Tortola and J.W.F.~Valle, \emph{{Are solar neutrino
  oscillations robust?}},
  \href{https://doi.org/10.1088/1126-6708/2006/10/008}{\emph{JHEP} {\bfseries
  10} (2006) 008} [\href{https://arxiv.org/abs/hep-ph/0406280}{{\ttfamily
  hep-ph/0406280}}].

\bibitem{Bakhti:2014pva}
P.~Bakhti and Y.~Farzan, \emph{{Shedding light on LMA-Dark solar neutrino
  solution by medium baseline reactor experiments: JUNO and RENO-50}},
  \href{https://doi.org/10.1007/JHEP07(2014)064}{\emph{JHEP} {\bfseries 07}
  (2014) 064} [\href{https://arxiv.org/abs/1403.0744}{{\ttfamily 1403.0744}}].

\bibitem{Coloma:2016gei}
P.~Coloma and T.~Schwetz, \emph{{Generalized mass ordering degeneracy in
  neutrino oscillation experiments}},
  \href{https://doi.org/10.1103/PhysRevD.94.055005}{\emph{Phys. Rev. D}
  {\bfseries 94} (2016) 055005}
  [\href{https://arxiv.org/abs/1604.05772}{{\ttfamily 1604.05772}}].

\bibitem{Colaresi:2021kus}
J.~Colaresi, J.I.~Collar, T.W.~Hossbach, A.R.L.~Kavner, C.M.~Lewis,
  A.E.~Robinson et~al., \emph{{First results from a search for coherent elastic
  neutrino-nucleus scattering at a reactor site}},
  \href{https://doi.org/10.1103/PhysRevD.104.072003}{\emph{Phys. Rev. D}
  {\bfseries 104} (2021) 072003}
  [\href{https://arxiv.org/abs/2108.02880}{{\ttfamily 2108.02880}}].

\bibitem{Coloma:2024ict}
P.~Coloma, E.~Fern\'andez-Mart\'\i{}nez, J.~L\'opez-Pav\'on, X.~Marcano,
  D.~Naredo-Tuero and S.~Urrea, \emph{{Improving the Global SMEFT Picture with
  Bounds on Neutrino NSI}},  \href{https://arxiv.org/abs/2411.00090}{{\ttfamily
  2411.00090}}.

\bibitem{SNO:2011hxd}
{\scshape SNO} collaboration, \emph{{Combined Analysis of all Three Phases of
  Solar Neutrino Data from the Sudbury Neutrino Observatory}},
  \href{https://doi.org/10.1103/PhysRevC.88.025501}{\emph{Phys. Rev. C}
  {\bfseries 88} (2013) 025501}
  [\href{https://arxiv.org/abs/1109.0763}{{\ttfamily 1109.0763}}].

\bibitem{Bahcall:1988em}
J.N.~Bahcall, K.~Kubodera and S.~Nozawa, \emph{{Neutral Current Reactions of
  Solar and Supernova Neutrinos on Deuterium}},
  \href{https://doi.org/10.1103/PhysRevD.38.1030}{\emph{Phys. Rev. D}
  {\bfseries 38} (1988) 1030}.

\bibitem{Chatterjee:2021wac}
S.S.~Chatterjee, P.S.B.~Dev and P.A.N.~Machado, \emph{{Impact of improved
  energy resolution on DUNE sensitivity to neutrino non-standard
  interactions}}, \href{https://doi.org/10.1007/JHEP08(2021)163}{\emph{JHEP}
  {\bfseries 08} (2021) 163}
  [\href{https://arxiv.org/abs/2106.04597}{{\ttfamily 2106.04597}}].

\bibitem{Rodrigues:2016xjj}
P.~Rodrigues, C.~Wilkinson and K.~McFarland, \emph{{Constraining the GENIE
  model of neutrino-induced single pion production using reanalyzed bubble
  chamber data}},
  \href{https://doi.org/10.1140/epjc/s10052-016-4314-3}{\emph{Eur. Phys. J. C}
  {\bfseries 76} (2016) 474}
  [\href{https://arxiv.org/abs/1601.01888}{{\ttfamily 1601.01888}}].

\end{thebibliography}\endgroup

\end{document}